\documentclass[sigconf, nonacm]{acmart}

\usepackage[ruled, boxed,vlined,linesnumbered]{algorithm2e}
\usepackage{algorithmic}
\usepackage{caption}
\usepackage{graphicx}
\usepackage{amsmath}
\usepackage{subfig} 
\usepackage{enumitem} 
\usepackage{multirow} 
\usepackage[toc,page]{appendix}
\usepackage{array}
\usepackage{tikz} 
\usepackage{threeparttable} 
\usepackage{cancel}

\newcommand{\ignore}[1]{}

\newcommand{\namesys}{\textit{HAKES}}
\newcommand{\nameindex}{\textit{HAKES-Index}}

\DeclareMathOperator*{\argmin}{arg\,min}
\DeclareMathOperator{\softmax}{softmax}

\begin{document}
\title{\namesys{}: Scalable Vector Database for Embedding Search Service}

\author{Guoyu Hu$^{1}$, Shaofeng Cai$^1$, Tien Tuan Anh Dinh$^2$, Zhongle Xie$^3$,}
\author{Cong Yue$^1$, Gang Chen$^3$, Beng Chin Ooi$^{1,3}$}
\affiliation{%
  \institution{$^1$National University of Singapore $^2$Deakin University  $^3$ 
  Zhejiang University }
  }
\email{{guoyu.hu, shaofeng, yuecong, ooibc}@comp.nus.edu.sg, anh.dinh@deakin.edu.au, {xiezl, cg}@zju.edu.cn }

\begin{abstract}
Modern deep learning models capture the semantics of complex data by transforming them into high-dimensional embedding vectors.
Emerging applications, such as retrieval-augmented generation, use approximate nearest neighbor (ANN) search in the embedding vector space to find similar data.
Existing vector databases provide indexes for efficient ANN searches, with graph-based indexes being the most popular due to their low latency and high recall in real-world high-dimensional datasets.
However, these indexes are costly to build, suffer from significant contention under concurrent read-write workloads, and scale poorly to multiple servers. 

Our goal is to build a vector database that achieves high throughput and high recall under concurrent read-write workloads. 
To this end, we first propose an ANN index with an explicit two-stage design combining a fast filter stage with highly compressed vectors and a refine stage to ensure recall, and we devise a novel lightweight machine learning technique to fine-tune the index parameters.
We introduce an early termination check to dynamically adapt the search process for each query. 
Next, we add support for writes while maintaining search performance by decoupling the management of the learned parameters. 
Finally, we design HAKES, a distributed vector database that serves the new index in a disaggregated architecture. 
We evaluate our index and system against 12 state-of-the-art indexes and three distributed vector databases, using high-dimensional embedding datasets generated by deep learning models. 
The experimental results show that our index outperforms index baselines in the high recall region and under concurrent read-write workloads. 
Furthermore, \namesys{} is scalable and achieves up to $16\times$ higher throughputs than the baselines. The HAKES project is open-sourced~\footnote{\url{https://www.comp.nus.edu.sg/~dbsystem/hakes/}}.
\end{abstract}

\maketitle

\section{introduction}
\label{sec:intro}
High-dimensional embedding vectors generated by deep learning models are becoming an important form of data
representation for complex, unstructured data such as images~\cite{clip, embedding-search-image}, audios~\cite{embedding-audio-1}, 
and texts~\cite{dpr, e5}.
The models convert input data to vectors in an embedding space and capture the data semantics relevance by heir relative positions in the high-dimensional space. 
Typical embedding vectors nowadays have hundreds to thousands of dimensions.

\begin{figure}
  \centering
  \subfloat{\includegraphics[width=0.475\textwidth]{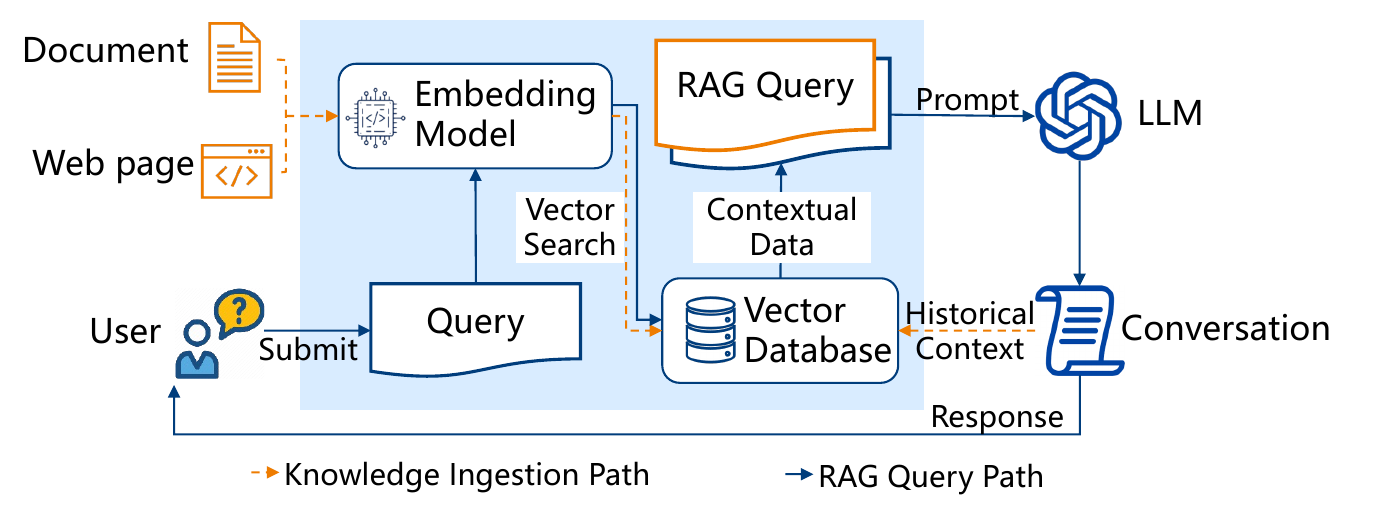}}
  \vspace{-0.3cm}
  \caption{Vector database in retrieval augmented generation.}
  \Description[short description]{long description}
  \label{fig:rag}
\end{figure}

Vector databases are designed to support efficient nearest neighbor search in the vector space.  They underlie many modern applications, ranging from search engines
~\cite{lanns, embedding-search-image}
, recommendation systems~\cite{recommend-product} to retrieval-augmented generation (RAG)~\cite{rag-ref-1, rag-ref-2}
These applications require efficient, high quality search as well as support for database updates.  
Figure~\ref{fig:rag} shows an example of how a vector database is used in RAG applications. 
A user submits a query to RAG, which turns the query into a vector. In the next step, RAG performs nearest neighbor search to find semantically similar data stored in a vector database. It then augments the query with the data found in the previous step and sends the new query to an LLM. To improve future responses, RAG frequently updates the vector database with data representing new knowledge, such as new documents, web pages, and past user interactions.

Vector databases use indexes to support efficient nearest neighbor search. Since searching for exact nearest neighbors
is too costly due to the curse of dimensionality~\cite{curse-of-dim}, existing works on vector indexes focus
on approximate nearest neighbor (ANN) search. 
The vast number of proposed ANN indexes can be classified as
graph-based or partitioning-based indexes~\cite{hnsw, faiss, lvq, adsampling, bench-tkde19, scann}.
These indexes are mostly evaluated using read-only workloads. Many of the datasets used for evaluation, such as Deep, Sift, and Glove, have lower dimensions than the deep embedding vectors used in emerging applications~\cite{hnsw, ann-benchmarks, scann, lccs-lsh}.
We identify three limitations of vector databases built around the existing ANN indexes to support modern applications, for which graph-based indexes are the recommended choice.
 
The first limitation is the computation overhead under high-dimensional spaces. In particular, comparing a vector against
its neighbors becomes more expensive with higher dimensions. Graph-based indexes~\cite{nsg, hnsw, diskann, bench-tkde19} are very costly to build because they require connecting each data point to its near neighbors and optimizing the graph structure to enable efficient traversal. 
The second limitation is the search performance under concurrent read-write workloads. Updating an existing index can be done in-place~\cite{hnsw, lsh-apg,
spfresh}, or out-of-place using a separate data structure and performing periodic consolidation~\cite{milvus, cassandra-5.0, analyticdb-v}. Graph indexes perform in-place updates, and require fine-grained locking over the neighborhood of the nodes on its traversal path~\cite{hnsw, milvus}. This results in significant
read-write contention. Out-of-place updates, on the other hand, require a separate search on the newly inserted data, while only postponing the update cost to a later time.
The third limitation is scalability. Existing vector databases treat their indexes as black boxes~\cite{lanns,manu,analyticdb-v}. 
They shard the data and build an independent graph index for each shard. 
However, to achieve high recall in high-dimensional spaces, nearly all data shards are searched.
The large number of searches per query leads to low throughput.

We present \namesys{}, a scalable vector database that achieves high recall and throughput under concurrent read-write workloads.
The database adopts a filter-and-refine design that consists of two stages. 
The filter stage narrows down the search candidates using compressed vectors for efficiency. 
The refine stage ranks the candidates based on the full-precision vectors.
The system addresses the first limitation by employing dimensionality reduction, coarse-grained partitioning, and quantization techniques. Furthermore, it proposes a novel light-weight machine learning technique to optimize the index parameters such that the filter stage is efficient and returns a set of high-quality candidates. 
\namesys{} also includes an early termination check at the filter stage to avoid unnecessary processing.
The compressed vectors are grouped by IVF index in contiguous buffers, and decoupling the index parameters used for compressing the vectors and those used during search enables seamless integration of new vectors at minimal overhead and contention, addressing the second limitation.
\namesys{} addresses the scalability limitation by exploiting the decoupling of the filter and refine stage to deploy them in a disaggregated architecture.
It distributes the memory and computation cost over multiple nodes, thereby achieving high throughput at scale. 

\namesys{} combines and adapts known techniques in a novel way to achieve its goal. In particular, existing works on quantization aim to improve the quality of similarity score approximation over the compressed vectors, minimizing the need for reranking the full-precision vectors~\cite{lvq, scann, rabitq, spreading, quip}.
\namesys{} aims to achieve good throughput-recall tradeoffs overall. By having a separate refine stage that reranks the original vectors, the dimensionality reduction and quantization aim to compress the vectors aggressively to reduce the computation cost at the filter stage. The compression parameters are learned in an end-to-end manner, in which the objective is to minimize the similarity score distribution distortion locally for vectors close to each other. The learning approach in \namesys{} does not assume access to external information, such as the embedding generation models, ground truth neighbors, or semantic labels, which is a different problem setting compared to other works that employ learning to improve the retrieval quality~\cite{repconc, distill-vq}. Moreover, our system allows for applying the newly learned parameters during search directly without re-indexing vectors in the database. In other words, learning can be done asynchronously while the vector database serves queries. Finally, the early termination check in \nameindex{} is more lightweight than that in~\cite{laet, leqat}, and more effective in our context than those in~\cite{auncel, idistance, vbase}, since it does not rely on accurate similarity scores under compression.

In summary, we make the following contributions:
\begin{itemize}[leftmargin=*]    
    \item We propose a novel index, \nameindex{}, that combines a compressed partitioning-based index with dimensionality reduction and quantization. The index leverages a lightweight machine learning technique to generate high-quality candidate vectors, which are then refined by exact similarity computation. It allows terminating the search early based on the intermediate results.  
    
    \item We propose a technique that decouples index parameters for compressing vectors during updates from those used for similarity computation. This ensures high performance under concurrent read-write workloads.  
    
    \item We design a distributed vector database, called \namesys{}, employing the new index in a disaggregated architecture. The system achieves scalability by spreading out the memory and computation overhead over multiple nodes.

    \item We compare \nameindex{} and \namesys{} against 12 state-of-the-art indexes and three popular commercial distributed vector databases. We evaluate the indexes and systems using high-dimensional embedding vector datasets generated by deep
learning models.  The results demonstrate that \nameindex{} outperforms both partitioning-based and graph-based index
baselines.  Furthermore, \namesys{} is scalable, and achieves up to $16\times$ higher throughputs at high recall than
the three other baselines do. 
\end{itemize}

The remainder of the paper is structured as follows. 
Section~\ref{sec:background} provides the background on  ANN search and the state-of-the-art ANN indexes. 
Section~\ref{sec:index} and
Section~\ref{sec:system} describes the design of our index and the distributed vector database.  Section~\ref{sec:evaluation} evaluates our designs against state-of-the-art indexes and systems. Section~\ref{sec:related} reviews the related works, and Section~\ref{sec:conclusion} concludes.

\section{Preliminaries}
\label{sec:background}

\noindent{\bf Approximate nearest neighbor Search.}
Let $\mathcal{D}$ denote a dataset containing $N$ vectors in a $d$-dimensional vector space $\mathbb{R}^d$.
For a query vector $\mathbf{x}$,
the similarity between $\mathbf{x}$ and a vector $\mathbf{v} \in D$ is defined by a metric $d(\mathbf{x}, \mathbf{v})$.
Common metrics include the Euclidean distance, inner product, and cosine similarity.
A vector $\mathbf{v_i}$ is considered closer to $\mathbf{x}$ than $\mathbf{v_j}$ if $d(\mathbf{x}, \mathbf{v_i}) < d (\mathbf{x}, \mathbf{v_j})$.
The $k$ nearest neighbors of $\mathbf{x}$ are vectors in $\mathcal{R} \subseteq \mathcal{D}$, where   $|\mathcal{R}|=k$ and $\forall \mathbf{v} \in \mathcal{R}, \forall \mathbf{u} \in \mathcal{D} \backslash \mathcal{R}, d(\mathbf{x}, \mathbf{v}) \le d (\mathbf{x}, \mathbf{u})$.
Finding the exact set $\mathcal{R}$ in a high-dimensional space is expensive due to the curse of dimensionality~\cite{curse-of-dim}. Instead, existing works on vector databases focus on  approximate nearest neighbor (ANN) search, which use ANN indexes to quickly find a set $\mathcal{R}'$ of vectors that are close to, but not necessarily nearest to $x$.
The quality of $\mathcal{R}'$ is measured by its {\em recall} relative to the exact nearest neighbor set, computed as $\frac{|\mathcal{R} \cap \mathcal{R}'|}{|\mathcal{R}|}$. 
We discuss two major classes of ANN indexes below.

\begin{figure}[t]
  \centering
  \vspace{-0.3cm}
  \subfloat[Graph index\label{fig:graph-index-a}]{\includegraphics[width=0.184\textwidth]{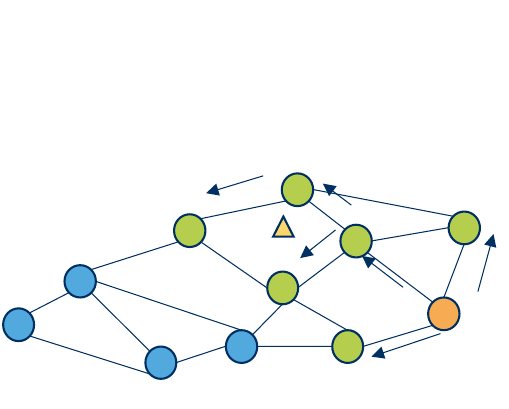}}
  \subfloat[HNSW \label{fig:graph-index-b}]{\includegraphics[width=0.178\textwidth]{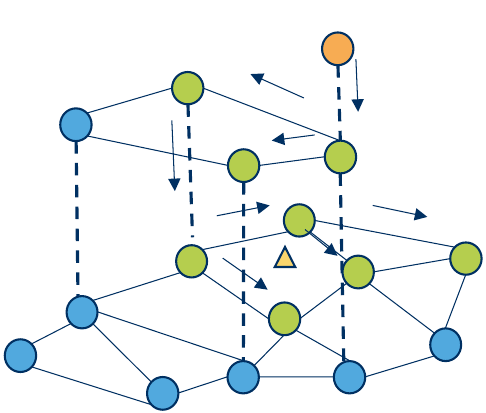}}
  \subfloat{\includegraphics[width=0.12\textwidth]{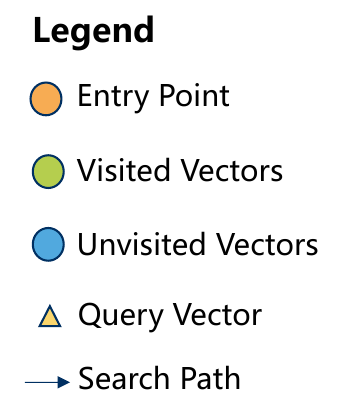}}
  \vspace{-0.3cm}
  \caption{Graph-based ANN index.}
  \Description[short description]{long description}
  \label{fig:graph-index}
\end{figure}

\noindent{\bf Graph-based indexes.} 
They build a proximity graph in which the vertices are the vectors, and an edge between two vertices means the two corresponding vectors are similar~\cite{nsg, hnsw, bench-tkde19}.
An ANN query involves a greedy beam search that starts from an entry point to locate close neighbors.
The query maintains a fixed-size set of candidates and visited nodes during the traversal.
At each step, the nearest unvisited vector from the candidate set is selected, and its unvisited neighbors are new potential candidates.
These new candidate vectors are evaluated for their similarity scores against the query vector and added to the candidate set accordingly.
The process repeats until the candidate set contains only visited nodes, as illustrated in Figure~\ref{fig:graph-index-a}.
When building or adding new vectors to the graph, a similar search is conducted to find the nodes to be connected based on a condition that allows future queries to reach their nearest neighbors and in a small number of steps~\cite{hnsw, diskann, ssg}. 
Since the search efficiency and recall depend on the graph, most existing works on graph indexes focus on building and maintaining a high-quality graph~\cite{hnsw, nsg, ssg, lsh-apg}.

The Hierarchical Navigable Small World graph (HNSW) is the most popular graph index.
It supports incremental updates and efficient search by introducing a
hierarchical structure 
with an exponentially decreasing number of vertices from the bottom to the top level, as shown in Figure~\ref{fig:graph-index-b}. 
A search starts from an entry point at the top level. At each level, it finds the nearest neighbor and starts the search in the next level with that vertex. Finally, at the bottom level, it performs beam search to find nearest neighbors.
During an update (i.e. adding a new vector), the new vertex's neighbors are first located at each level, and then the edges are updated. The update condition restricts the number of neighbors and only adds an edge if the similarity between the searched candidate and the new vector is larger than that of the new vector and its existing neighbors. 
This update process is costly, and it creates significant contention under concurrent read-write workloads.

\begin{figure}[t]
  \centering
  \vspace{-0.3cm}
  \subfloat[Index partitioning \label{fig:partition-index-a}]{\includegraphics[width=0.135\textwidth]{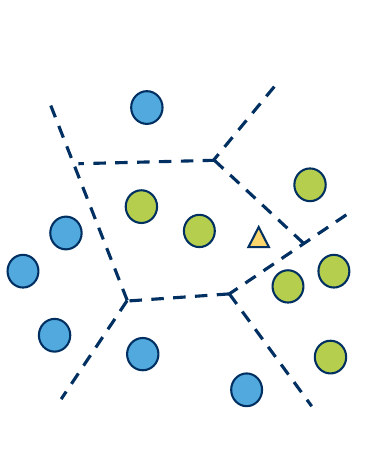}}
  \subfloat[PQ\label{fig:partition-index-b}]{\includegraphics[width=0.225\textwidth]{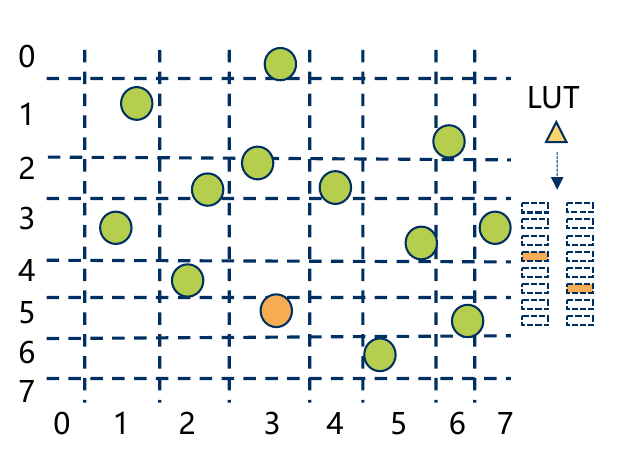}}
  \subfloat{\includegraphics[width=0.12\textwidth]{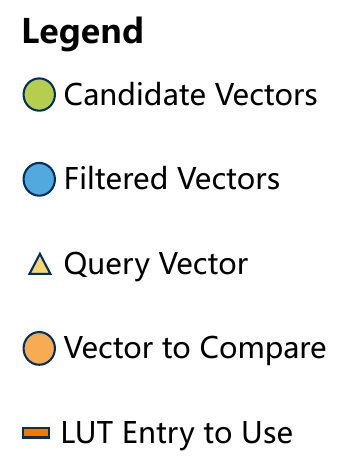}}
  \vspace{-0.3cm}
  \caption{Partitioning-based ANN index.}
  \Description[short description]{long description}
  \label{fig:partition-index}
\end{figure}

\noindent{{\bf Partitioning-based indexes.}}
They divide vectors into multiple partitions using one or multiple hashing schemes, such that similar vectors are in the same partition.
The similarity of a query vector to all the vectors in a partition can be approximated by its proximity to the partition itself. The partition assignments can be encoded for efficient search.
Examples of hashing schemes include locality sensitive hashing (LSH)~\cite{sh, lccs-lsh, falconnlsf}, clustering~\cite{idistance, faiss}, quantization~\cite{pq, ivfpqr, opq}, or neural networks~\cite{neurallsh, bliss,batl}. 
New vectors are added to the corresponding partition by computing its partition assignment. 
{A search for vector $\mathbf{x}$ starts by finding the partitions closest to $\mathbf{x}$, then retrieving the vectors belonging to the selected partitions, as shown in Figure~\ref{fig:partition-index-a}. Finally the $k$ closest vectors are selected by evaluating the similarity scores.}

Inverted-file (IVF) and product quantization are the most popular partitioning-based indexes. IVF~\cite{faiss, pq} uses k-means to partition the vectors. Specifically, a sample set of vectors is used to determine the cluster centroids, and then vectors belonging to the closest centroids are stored together in respective buckets. 
During a search, all partitions are ranked based on the similarity between their centroids and the query vector $\mathbf{x}$. The top $nprobe$ partitions are scanned to produce $k$ nearest neighbors. The number of centroids $N_c$ for k-means and the $nprobe$ determine the cost of ranking partitions and the number of candidate vectors. These parameters also affect recalls. For example, for million-scale datasets, high recalls can be achieved when $N_c$ is in 1000s and $nprobe$ is in 10s to 100s.

Product quantization (PQ) splits the original $d$-dimensional space into $m$ orthogonal subspaces of the same dimension $d'=d/m$. 
Each subspace is further partitioned, e.g., using k-means with $N_c$ centroids, resulting in $(N_c)^m$ partitions.
A codebook $\mathbf{C^{PQ}} \in \mathbb{R} ^{N_c\times d}$
is the concatenation of subspace centroids $\mathbf{C^{PQ}}_j \in \mathbb{R} ^{N_c\times d'}$, i.e., $\mathbf{C^{PQ}} =
[\mathbf{C^{PQ}}_1,\mathbf{C^{PQ}}_2, ..., \mathbf{C^{PQ}}_m]$. 
{
A vector can be quantized into a concatenation of indexes of the centroids in the codebook at each subspace,
$p(v) = [p_1(\mathbf{v}),p_2(\mathbf{v}), ..., p_m(\mathbf{v})]$, 
where $p_j(\mathbf{v}) = \argmin_{i} || \mathbf{C^{PQ}}_j[i]-\mathbf{v}_j ||$ denotes the index of the closest centroid in the $j^{th}$ subspace centroids $\mathbf{C^{PQ}_j}$.
Let $q_j(\mathbf{v}) = \mathbf{C^{PQ}}_j[p_j(\mathbf{v})]$ be the closest centroid of $\mathbf{v}$.}
The concatenation of centroids closest to $\mathbf{v}$ in respective subspaces forms its approximation:
$\mathbf{v} \approx q(\mathbf{v}) = [q_1(\mathbf{v}), q_2(\mathbf{v}), ..., q_m(\mathbf{v})]$.
Then, the similarity between a vector $\mathbf{x}$ and a vector $\mathbf{v}$ can be approximated as $d(\mathbf{x}, q(\mathbf{v}))$. For the commonly used
Euclidean distance (normally without taking the square root) and inner product, we have:

\begin{equation}
    d(\mathbf{x}, \mathbf{v}) \approx d(\mathbf{x}, q(\mathbf{v})) = \sum_{j=1..m}{d(\mathbf{x}_j, q_j(\mathbf{v}))}
    \label{eq:pq-sum}.
\end{equation}
PQ enables efficient comparison of $\mathbf{x}$ against the candidate vectors.
During a search, the query vector is split into $m$ subvectors, each of which is compared against all the centroids in its corresponding subspace, $\mathbf{C^{PQ}}_j$.
The resulting similarity scores are stored in a lookup table, LUT $\in \mathbb{R}^{N_c \times m}$.
Given Equation~\ref{eq:pq-sum}, the similarity between $\mathbf{x}$ and any vector $\mathbf{v}$ can then be approximated using the quantized vector $q(\mathbf{v})$ via $m$ lookups into the LUT, followed by a summation, as shown in Figure~\ref{fig:partition-index-b}.
In practice, PQ generates compact vector representations.
Typically, $N_c$ is 16 and 256, such that only 4 or 8 bits can encode the vector in each subspace. 
Recent indexes using 4-bit PQ  yield a LUT small enough to fit in CPU caches, and optimized the quantized vector layout for efficient SIMD implementation, thereby achieving significantly higher throughputs~\cite{fastscan, faiss, scann}. In practice, quantization is often used together with a coarse-grained partitioning technique such as IVF 
to filter out a large number of vectors before applying quantization.  Furthermore, an IVF partition stores vectors in a contiguous memory region, resulting in fast scanning of the quantization codes due to memory prefetching.
Quantization can also be used with graph indexes to speed up comparison of the query vector with the graph vertices~\cite{diskann, lvq}. Specifically, during traversal, comparison against each vertex is based on the quantized vectors instead of the original ones.

Quantization enables efficient but lossy approximation of the similarity scores between vectors.
Reranking the candidates can be performed after quantization to improve recall, using additional information~\cite{ivfpqr, lvq} or the original vectors~\cite{rabitq, faiss, scann}.
Some existing works on quantization, namely~\cite{scann, lvq, rabitq, quip},  focus on reducing the approximation errors in order to minimize reranking. Others aim to transform the vectors to be more suitable for quantization~\cite{opq, spreading}, or leverage information about the downstream task and upstream embedding model to improve end-to-end retrieval quality~\cite{distill-vq, repconc}.

\section{HAKES-Index}
\label{sec:index}

\begin{figure*}[t]
  \centering
  \vspace{-0.3cm}

  \subfloat[\nameindex{}\label{fig:hakes-index-a}]{\includegraphics[width=0.282\textwidth]{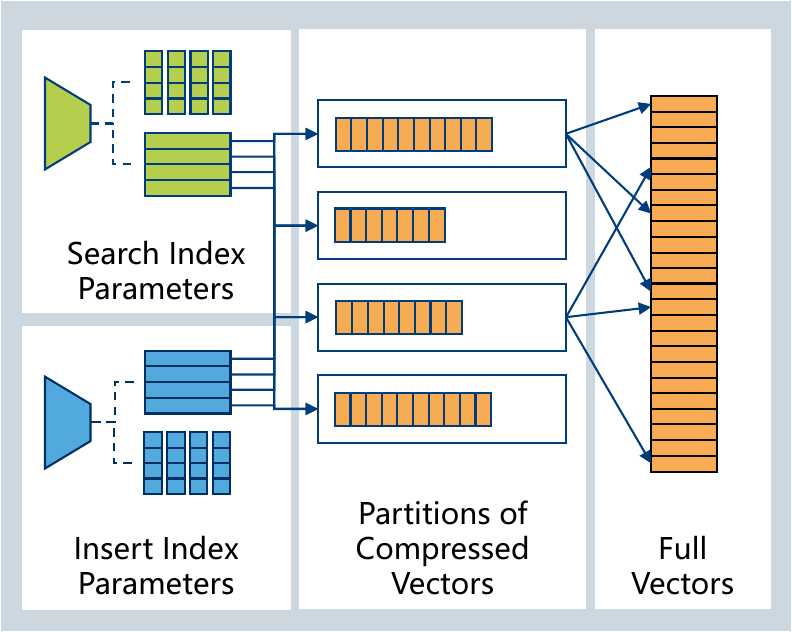}}
  \subfloat[Search\label{fig:hakes-index-b}]{\includegraphics[width=0.272\textwidth]{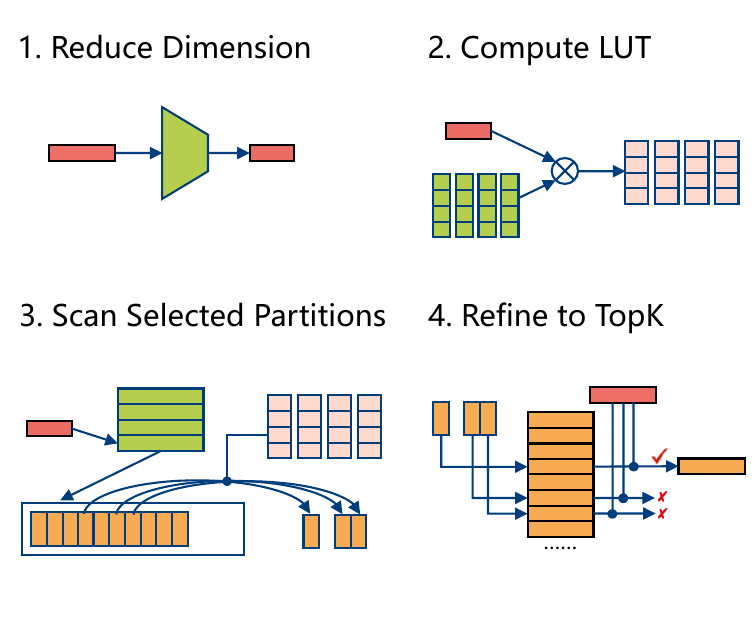}}
  \subfloat[Insert\label{fig:hakes-index-c}]{\includegraphics[width=0.26\textwidth]{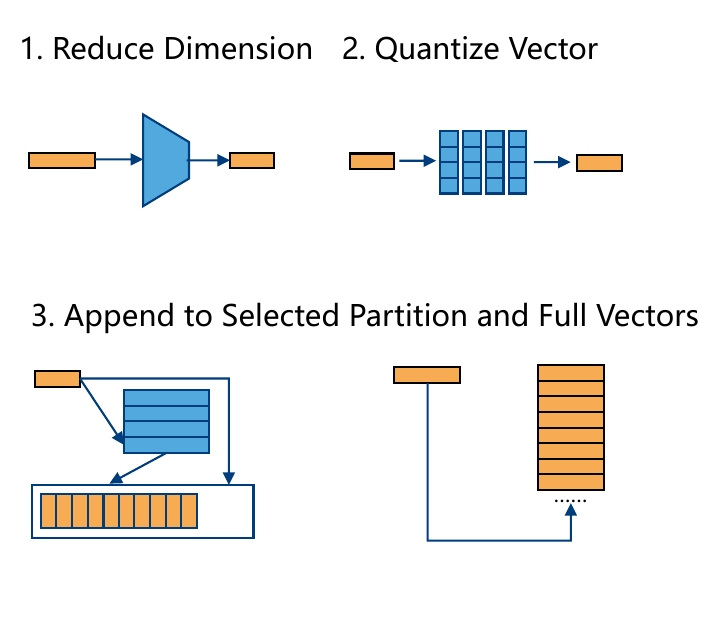}}
  \subfloat{\includegraphics[width=0.155\textwidth]{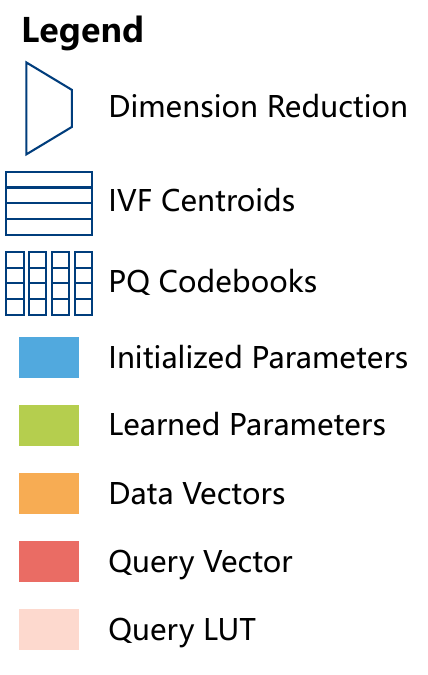}}
  
  \vspace{-0.3cm}
  \caption{\nameindex{} overview.}
  \vspace{-0.2cm}
  \Description[short description]{long description}
  \label{fig:hakes-index}
\end{figure*}

In this section, we present a novel index, called \nameindex{}, that supports efficient search and index update.

\subsection{Overview}
\label{subsec:index:overview}

Figure~\ref{fig:hakes-index-a} shows the components in \nameindex{}. 
It consists of three parts, the two respective sets of index parameters to process the vectors for search and insert, the partitions that contain compressed vectors, and the full vectors. 
Each set of index parameters is composed of a dimensionality reduction module, IVF centroids, and a PQ codebook.  
The compressed vectors are partitioned by the IVF centroids and the compression involves dimensionality reduction followed by quantization guided by the codebook.
The dimensionality reduction module uses a transformation matrix $\mathbf{A}\in \mathbb{R}^{d \times d_r}$ 
and a bias vector $\mathbf{b} \in \mathbb{R}^{d_r}$ to compress vectors from the original $d$-dimensional space to that of $d_r$-dimensional spaces, where $d_r < d$.
The IVF centroids, $\mathbf{C^{IVF}}$, determine the partition a new vector is attached to during the insert and rank the partitions for a query vector during the search.
The quantization codebooks, $\mathbf{C^{PQ}}$, are used to generate the quantized vector stored in the partitions and compute the Lookup table for search. 
Note that dimensionality reduction is placed at the front, which speeds up all subsequent computations.
We use $\mathbf{A}, \mathbf{b}, \mathbf{C^{IVF}}, \mathbf{C^{PQ}}$ to refer to the insert index parameters and $\mathbf{A'}, \mathbf{b'}, \mathbf{C^{IVF'}}, \mathbf{C^{PQ'}}$ to the search index parameters.

Search query involves four steps, shown in Figure~\ref{fig:hakes-index-b}.
Step 1 reduces the dimensionality of the query vector from $d$ to $d_r$ with $\mathbf{A'}, \mathbf{b'}$.
Next, the output of the dimensionality reduction is used to compute the lookup table (LUT) with the quantization codebook $\mathbf{C^{PQ'}}$ in step 2.
Step 3 evaluates the $d_r$-dimension query vector with $\mathbf{C^{IVF'}}$ to select the closest partitions for scanning using the LUT. 
$k' > k $ candidates are selected, and step 4 obtains the top $k$ of them by comparing the query vector to their full vectors.
The four steps in the search workflow can be mapped into two stages. 
The filter stage spans steps 1-3, where the majority of vectors are filtered out, leaving $k'$ candidate vectors. 
The last step is the refine stage, when $k'$ candidates are refined to the top $k$ nearest vectors.

To add vectors to \nameindex{}, the insert index parameters are utilized, as shown in Figure~\ref{fig:hakes-index-c}.
Each new vector is transformed using the dimensionality reduction parameters (step 1) and quantized using the codebook (step 2).
It is then appended to both the corresponding partition determined by the IVF centroids and the buffer holding full vectors.
For deletion, ~\nameindex{} uses tombstones to mark the deleted vectors. 
During the filter stage, the tombstones are checked before adding the vectors to the candidate set.
The deleted vectors and their corresponding compressed vectors are removed by a compaction step that rewrites the partitions. This step happens when the index is being checkpointed or rebuilt. The latter is triggered by an update in the embedding model, or when the data size grows beyond certain sizes. 
This approach reduces the interference of deletion on the search and insert operations.

\subsection{Index Construction}
\label{subsec:index:build}

\begin{figure*}[t]
  \centering
  \includegraphics[width=0.98\textwidth]{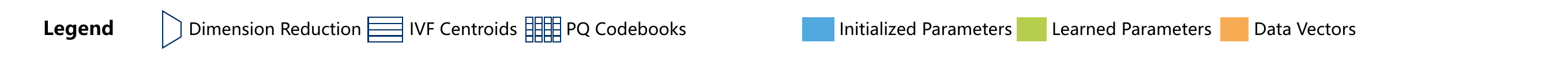}\\
  \vspace{-0.5cm}
  \begin{minipage}[b]{0.70\textwidth}
  \subfloat[Build base index\label{fig:build-a}]{\includegraphics[width=0.258\textwidth]{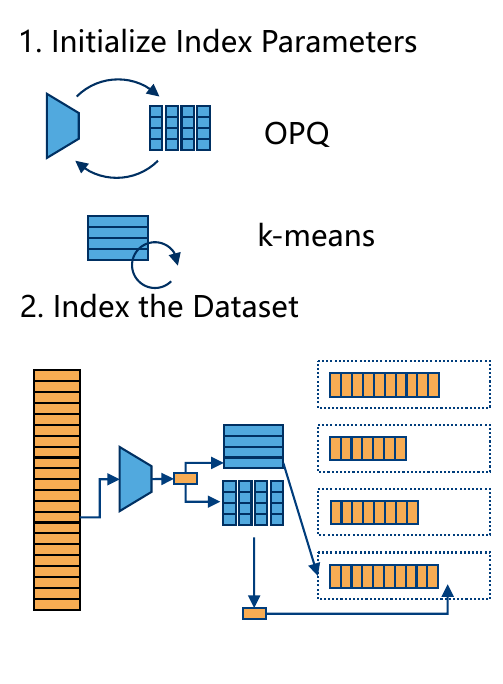}}
  \subfloat[Prepare training data\label{fig:build-b}]{\includegraphics[width=0.288\textwidth]{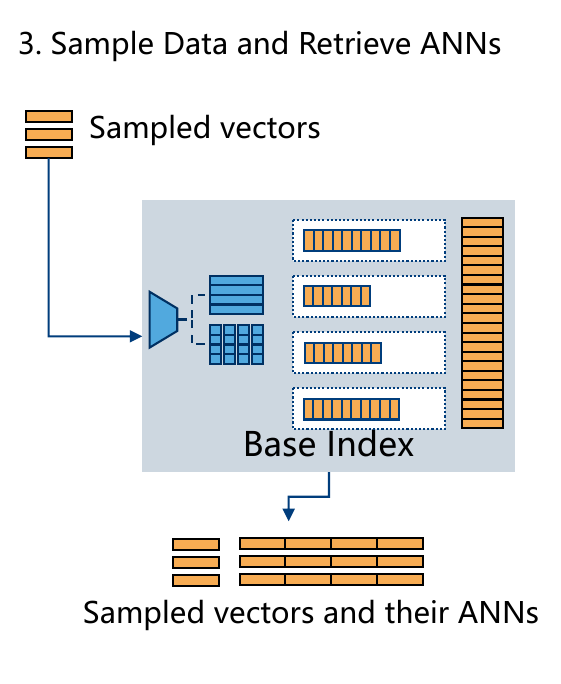}}
  \subfloat[Learn compression parameters\label{fig:build-c}]{\includegraphics[width=0.448\textwidth]{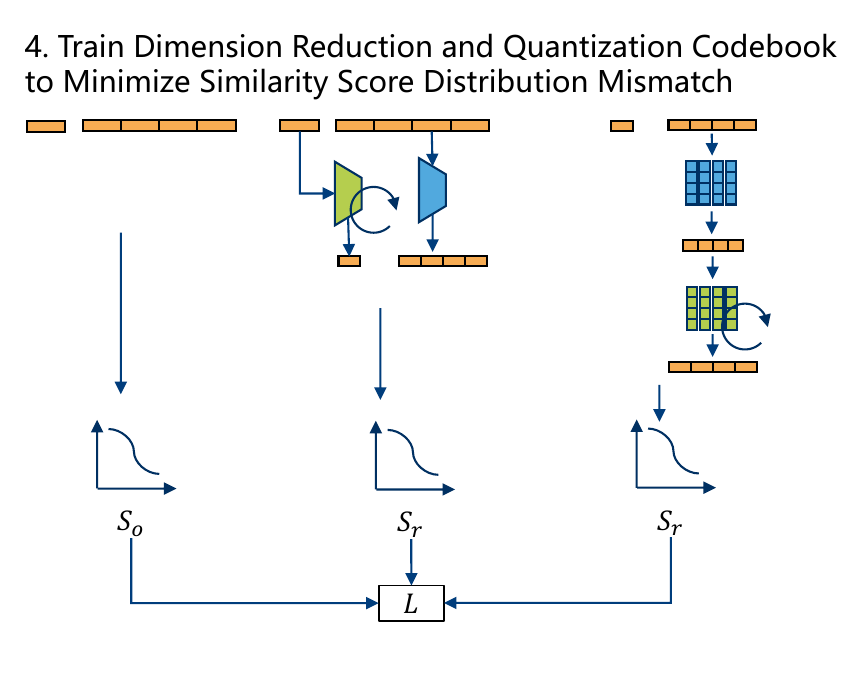}}
  \end{minipage}
  \begin{minipage}[b]{0.279\textwidth}
  \subfloat[Recalculate IVF centroids\label{fig:build-d}]{\includegraphics[width=0.99\textwidth]{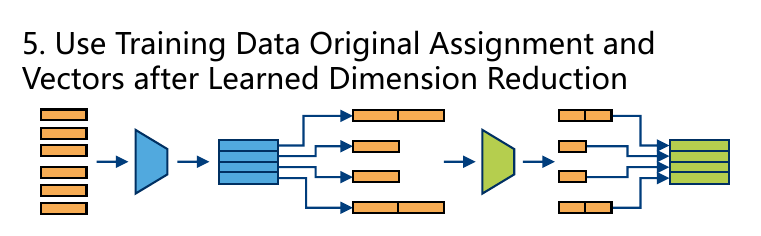}}
  \vspace{-0.2cm}
  \\
  \subfloat[Update index\label{fig:build-e}]{\includegraphics[width=0.99\textwidth]{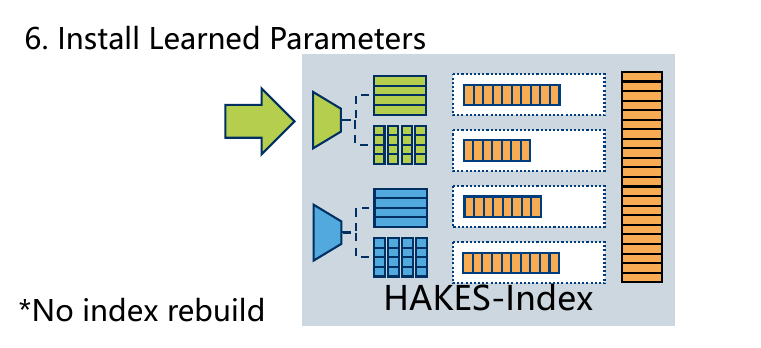}}
  \end{minipage}
  \caption{End-to-end index construction.}
  \Description[short description]{long description}
  \vspace{-0.2cm}
  \label{fig:build}
\end{figure*}

We construct \nameindex{} following the procedure illustrated in Figure~\ref{fig:build}.
Figure~\ref{fig:build-a} shows the first step of building the base index.
The insert index parameters are initialized with existing processes and then the dataset is inserted into the index.
Particularly, Optimal Product Quantization (OPQ) is employed to initialize $\mathbf{A}$ and $\mathbf{C^{PQ}}$, which iteratively finds a transformation matrix that minimizes the reconstruction error of a PQ codebook, and K-means is employed to initialize the IVF centroids, $\mathbf{C^{IVF}}$. The bias vector $\mathbf{b}$ is zero.
Next, the training set is prepared by sampling a set of vectors and obtaining their neighbors with the base index, as in Figure~\ref{fig:build-b}.
Note that another set of sampled pairs is used for validation.
Then, we use a self-supervised training method to learn the search parameters, $\mathbf{A'}$, $\mathbf{b'}$ and $\mathbf{C^{PQ'}}$, illustrated in Figure~\ref{fig:build-c}, which is the key to \nameindex{}'s high performance at high recall, and the technical details will be revealed in the Section~\ref{subsec:index:learn}.
After training, the IVF centroids $\mathbf{C^{IVF'}}$ are computed by partitioning the sample data with $\mathbf{A}, \mathbf{C^{IVF}}$, then recomputing the centroid for each partition after applying the learned $\mathbf{A'}$ and $\mathbf{b'}$ to vectors in it (Figure~\ref{fig:build-d}).
Finally, the newly learned $\mathbf{A'}$, $\mathbf{b'}$, $\mathbf{C^{PQ'}}$, and $\mathbf{C^{IVF'}}$, are installed in the index, as shown completed in 
Figure~\ref{fig:build-e}, serving subsequent search queries. 

The training process can run independently from the serving system.
In practice, the index is first built and uses $\mathbf{A}, \mathbf{b}, \mathbf{C^{IVF}}, \mathbf{C^{PQ}}$ for both insert and search. As it serves requests, the system records samples, and the training process runs in the background.
Once the training is finished, the new parameters $\mathbf{A'}, \mathbf{b'}, \mathbf{C^{IVF'}}, \mathbf{C^{PQ'}}$ can be used immediately to serve queries.
In other words, \nameindex{} can be updated incrementally. 
Moreover, the construction of \nameindex{} is efficient. That reduces the time to rebuild the index for serving at an updated throughput-recall frontier, when the database sizes and distributions are significantly changed by insertion and deletion.

\subsection{Learning Compression Parameters}
\label{subsec:index:learn}
Since the search recall depends on the quality of the candidate vectors returned by the filter stage, \nameindex{} achieves high recall by ensuring that there are many true nearest neighbors in the set of $k'$ candidate vectors. The {\em compression parameters} in \nameindex{}, which include $\mathbf{A'}$ and $\mathbf{b'}$ for dimensionality reduction, and $\mathbf{C^{PQ'}}$ for the PQ codebooks, are fine-tuned, so that they capture the similarity relationship between the query vector and indexed vectors.

At the beginning of training process, $\mathbf{A'}$ and $\mathbf{C^{PQ'}}$ are initialized with $\mathbf{A}$ and $\mathbf{C^{PQ}}$ that are produced by OPQ.
The bias vector $\mathbf{b'}$ is initialized with zero. 
We then jointly optimize $\mathbf{A'}$, $\mathbf{b'}$, and $\mathbf{C^{PQ'}}$ to minimize the mismatch between the similarity score distribution after quantization and that of the original $d$-dimensional space.
We only focus on the mismatch in a local region that the training objective is defined based on the similarity score distributions of a sampled query vector $\mathbf{x}$ and its close neighbors $ANN_x$, because distant vectors are filtered away by coarse grained IVF partition selection during search in \nameindex{}. 
Specifically, the similarity score distributions before and after the dimensionality reduction are:

\begin{equation}
\label{eq:score-dist-original}
    S_{\mathbf{o},\mathbf{x}} = \softmax([d(\mathbf{x}, \mathbf{v_1}), \dots, d(\mathbf{x}, \mathbf{v_K})])
\end{equation}

\begin{equation}
    \label{eq:score-dist-dr}
    S_{\mathbf{r},\mathbf{x}} = \softmax([d({R'}(\mathbf{x}), R(\mathbf{v_1})), \dots, d({R'}(\mathbf{x}), R(\mathbf{v_K}))])
\end{equation}

\noindent
where $K =  |ANN_x|$ is the number of retrieved close neighbors, and the softmax function converts the similarity scores to a distribution. $R'(\mathbf{x}) = \mathbf{A'x}+\mathbf{b'}$ and $R(\mathbf{v}) = \mathbf{Av}+\mathbf{b}$ represent dimensionality reduction.
The distribution of the similarity scores after quantization is:

\begin{equation}
\label{eq:score-dist-pq}
    S_{q,\mathbf{x}} = \softmax([d({R'}(\mathbf{x}), q'(R(\mathbf{v_1})), \dots, d({R'}(\mathbf{x}), q'(R(\mathbf{v_K}))])
\end{equation}

\noindent 
where the vector approximation $q'(\mathbf{v}) = [q'_1(\mathbf{v}), q'_2(\mathbf{v}), \dots, q'_m(\mathbf{v})]$ from PQ is modified to use both $C^{PQ}$ and $C^{PQ'}$. Specifically,
$ q_j(\mathbf{v}) = \mathbf{C^{PQ'}}_j[\argmin_{i} || \mathbf{C^{PQ}}_j[i]-\mathbf{v}_j ||]$. It means that the indexes of the centroids of the codebook are produced by $C^{PQ}$ and the fine-tuned centroids of $C^{PQ'}$ at the corresponding position are used to approximate the vector.

With the distributions of similarity scores, we can then reduce the mismatch by minimizing the Kullback-Leibler (KL) divergence defined over two pairs of distributions. One pair is defined between the distribution in the original vector space (Equation~\ref{eq:score-dist-original}) and that in the vector space after dimensionality reduction (Equation~\ref{eq:score-dist-dr}).
The other pair is between (Equation~\ref{eq:score-dist-original}) and the distribution of similarity scores calculated between a query vector after dimensionality reduction and its quantized close neighbors (Equation~\ref{eq:score-dist-pq}).
The overall training objective is as follows:

\begin{equation}
\label{eq:loss}
    L = -\sum_{\mathbf{x} \in D_{sample}} {S_o\log\frac{S_{r,\mathbf{x}}}{S_{o, \mathbf{x}}}} - \lambda \sum_{\mathbf{x} \in D_{sample}} {S_{o, \mathbf{x}}\log\frac{S_{q, \mathbf{x}}}{S_{o, \mathbf{x}}}}
\end{equation}

\noindent
where $D_{sample}$ is the sampled query vectors for training, and $\lambda$ is a hyperparameter to control the strength of the regularization.

The training process iteratively updates $\mathbf{A'}, \mathbf{b'}, \mathbf{C^{PQ'}}$ to minimize the loss defined in Equation~\ref{eq:loss} that is to minimize the mismatch among three similarity distributions for close vectors as illustrated in Figure~\ref{fig:build-c}. It stops when the loss reduction computed on the validation set is smaller than a threshold (e.g., 0.1).

\subsection{Search Optimizations}
\label{subsec:index:optim}
\nameindex{} contains two additional optimizations that improve search efficiency. The first is INT8 scalar quantization at each dimension of the IVF centroids. This allows using SIMD to evaluate $4\times$ more dimensions in a single instruction. Although quantization can be lossy, such representation errors are tolerable in practice since the centroids are only used for partition assignment and a large number of partitions are selected for high recall. 
The second optimization is to adapt the cost of the filter stage based on the query. Fixing the value of $nprobe$ means that the computation cost is roughly the same for every query. We note that in extreme cases, all the true nearest neighbors are in the same partition, where only that partition needs to be scanned. In other extreme cases, the true nearest neighbors are evenly distributed among the partitions, where all the partitions need to be scanned to achieve high recall. 
In high-dimensional space, it is challenging to determine the $nprobe$ based solely on the centroids. \nameindex{} introduces a heuristic condition for early stopping the scanning of subsequent partitions based on the intermediate search results. 
The search process ranks the partitions by the similarity score of their centroids to the query, and scans the partitions in order. The key idea is that, as the search process moves away from the query vector, new partitions will contribute fewer vectors to the candidate set.
We track the count of consecutively scanned partitions that each partition adds fewer than $t$ vectors to the candidate set, where $t$ is a search configuration parameter.
When that count exceeds a specified threshold $n_t$, it indicates that the search has likely covered all partitions containing nearest neighbors, and we terminate the filter stage.
\nameindex{} terminates the filter stage either when the heuristic condition above is met, or when $nprobes$ partitions have been scanned. 

\subsection{Discussion}
\label{subsec:index:discuss}

\nameindex{}'s two-stage design allows the filter stage to trade accuracy of similarity score evaluation for lower computation overhead.
This stage performs aggressive compression, combining dimensionality reduction at the beginning and then 4-bit product quantization.
The index parameters are optimized to achieve high compression ratios while preserving only the distribution and not the exact values of similarity scores. The optimization focuses on the local regions instead of globally, since distant vectors in IVF are already filtered out and never evaluated.
Our experimental results demonstrate that deep embedding vectors can be aggressively compressed to achieve superior throughput-recall tradeoff overall for \nameindex{}, with $d_r$ as small as ${1/4}$ or ${1/8}$ of the original dimension $d$, and with 4-bit PQ with $m=2$. The early termination checking is designed to operate in the filter stage with aggressive compression. It does not rely on accurate similarity score calculation, unlike existing works~\cite{auncel, idistance, vbase}. The statistics tracking and check incurs minimal overhead, compared to other works on early termination~\cite{laet, leqat}.

The compression techniques in \nameindex{} differs from those of existing works on quantization, which either focuses on minimizing the reconstruction error~\cite{pq, aq, lvq}, i.e., $d(\mathbf{v}, q(\mathbf{v}))$, or the error of similarity score approximation~\cite{quip, scann, soar}, i.e., $d(\mathbf{x}, q(\mathbf{v}))$. 
\nameindex{} learns both dimensionality reduction and quantization together to reduce the distortion of the similarity distribution. Some learned data transformations for quantization~\cite{opq, dopq} aim to transform the original vector to reduce the quantization error. ~\cite{spreading} introduces complex data transformation, increases serving complexity, and some other works tune even the embedding models~\cite{repconc, distill-vq}, which differ from our ultimate goal of achieving superior throughput-recall trade-off for the ANN search with given embedding vectors. Moreover, we only use the approximate nearest neighbor for training, which can be efficiently obtained compared to ground truth neighbors required by other works~\cite{spreading, repconc}.

A key design in \nameindex{} is that it decouples the management of parameters used for search and insert, enabling its high-recall search while supporting the incorporation of new data.
Specifically, it maintains two sets of compression parameters: the learned parameters obtained through training as the search index parameters, and the original parameters established upon initialization as the insert index parameters, as shown in Figure~\ref{fig:hakes-index-a}.
It is closely related to the lightweight self-supervised training process.
As discussed in Section~\ref{subsec:index:learn}, we use the prebuilt base index and fix PQ code assignment for training, where all the data vectors are processed only once using the original set of parameters.
Consequently, new vectors can follow the same process of being indexed by the initialized parameters and searched by the learned parameters.
Empirical observations also confirm that using the learned parameters for inserting new vectors leads to recall degradation in Section~\ref{sec:evaluation}.
Furthermore, as a consequence of the decoupling, the learned search index parameters can be directly applied without re-indexing the vectors.
Existing works on learned compression use the updated codebook for assignment during every training iteration~\cite{repconc, distill-vq}. They would require expensive re-indexing of the vectors when applying the trained parameters in vector databases to serve queries. 

The aggressive compression employed by \nameindex{} not only significantly speeds up the filter stage, but also reduces the memory consumption in this stage.
We now analyze the memory cost of \nameindex{} for a vector dataset of $(N\cdot4\cdot d)$ bytes. The dimensionality reduction matrices and the bias vector take $(2\cdot 4\cdot d\cdot d_r + 4 \cdot d_r)$ bytes. IVF centroids and the 4-bit quantization codebooks consume $(N_c\cdot 4 \cdot d_r + N_c \cdot d_r)$ bytes and $(2\cdot 2^4\cdot 4 \cdot d_r)$ bytes respectively. The compressed vectors take $(N\cdot (1/2) \cdot (d_r/m))$ bytes. 
The filter stage index is significantly smaller than the vector dataset.

As the dataset grows considerably, the index should be rebuilt with a larger number of IVF partitions. In practice, even the embedding models that generate the vectors are frequently retrained, for example, on a daily basis in recommendation systems~\cite{embedding-recsys-retrain-daily}). After model training, rebuilding the index is necessary.

\section{The \namesys{} Distributed VectorDB}
\label{sec:system}

In this section, we present the design of our distributed vector database, named \namesys{}. 

\subsection{Overview}
\label{subsec:sys:arch}

\nameindex{} processes a search query in two stages, namely the filter and refine stage. These stages do not share data, and they have distinct resource requirements due to the types and amount of vectors being evaluated. 
Specifically, the memory consumption of the filter stage, accessing compressed vectors, is significantly lower than that of the refine stage, which accesses the original vectors. 
In addition, the filter stage has a much higher computation cost because it performs computation over a large number of vectors. 

\begin{figure}
  \centering
  \subfloat{\includegraphics[width=0.475\textwidth]{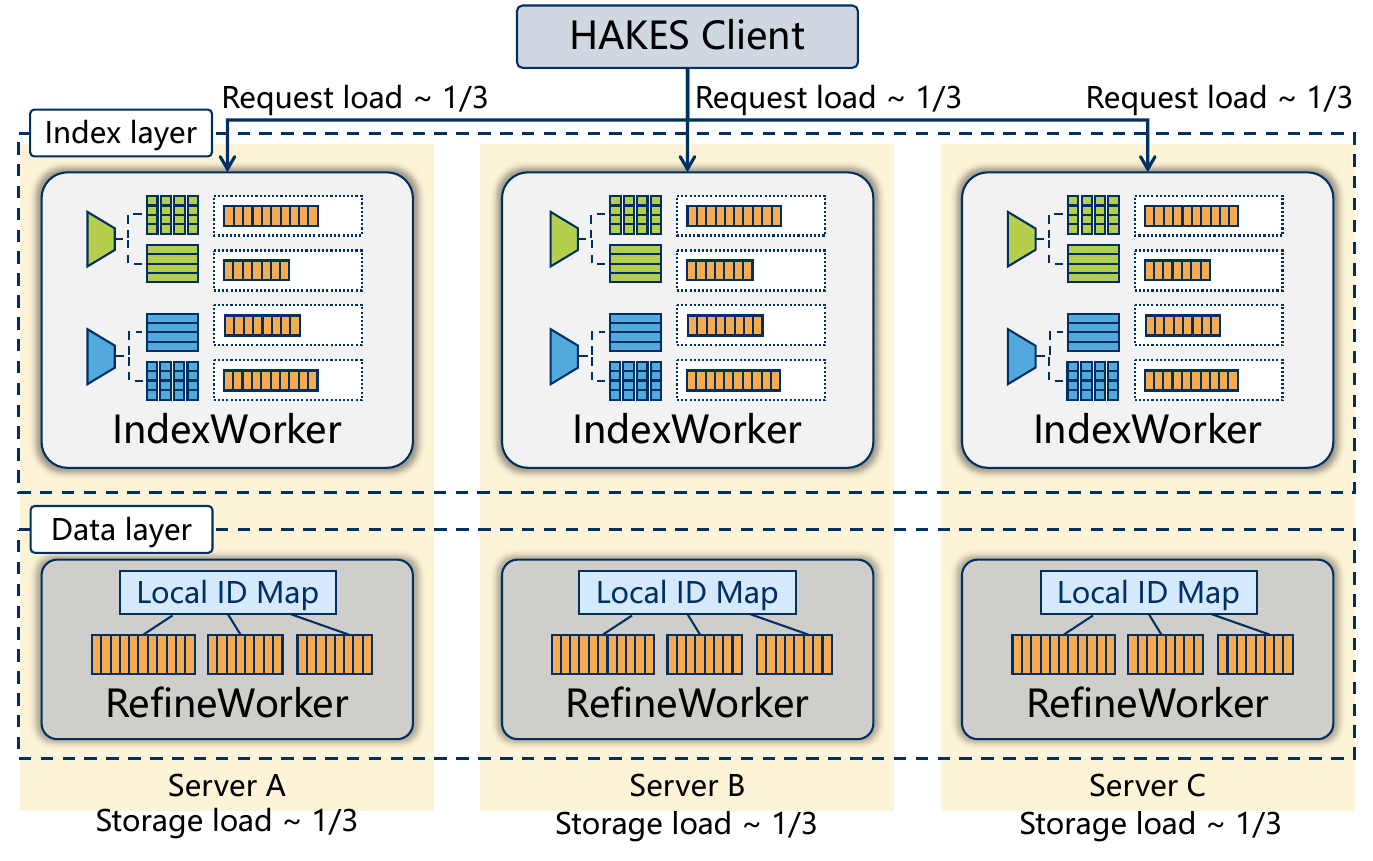}}
  \vspace{-0.3cm}
  \caption{\namesys{} architecture.}
  \Description[short description]{long description}
  \label{fig:arch}
\end{figure}

We design an architecture that exploits {the filter-and-refine design} to disaggregate the two stages. In particular, we separate the management of the filter-stage index from the full-precision, original vectors used only in the refine stage{, and employ different scaling policies for them in a server cluster. 
There are two sets of workers, the IndexWorkers and the RefineWorkers, each performs one stage of the index using the local data.}
The former are responsible for the filter stage, managing the replicated compressed vectors. 
The latter performs the refine stage, storing shards of the original vectors. 
Figure~\ref{fig:arch} shows an example in which a physical server runs both an IndexWorker and a RefineWorker. However, we stress that these components can be disaggregated and scaled independently. For example, more memory nodes running RefineWorkers can be added to handle a large volume of data, and more compute nodes running IndexWorker can be added to speed up the filter stage. 

\begin{figure}
  \centering
  \begin{minipage}[c]{0.35\textwidth}
  \subfloat[Design I\label{fig:typical-arch-a}]{
  \hspace{-.3cm}\includegraphics[width=.48\textwidth]{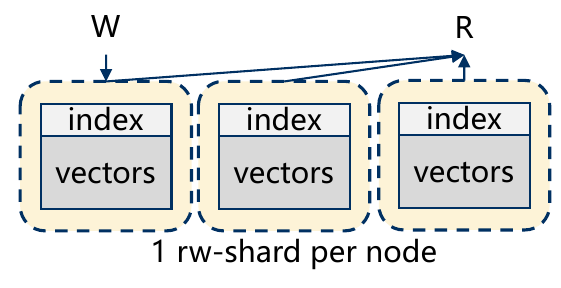}}
  \subfloat[Design II\label{fig:typical-arch-b}]{\includegraphics[width=.5\textwidth]{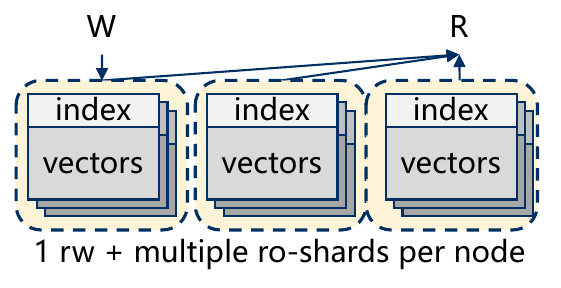}}
  \vspace{-.4cm}
  \\
  \subfloat[Design III\label{fig:typical-arch-c}]{
  \hspace{-.3cm}\includegraphics[width=.5\textwidth]{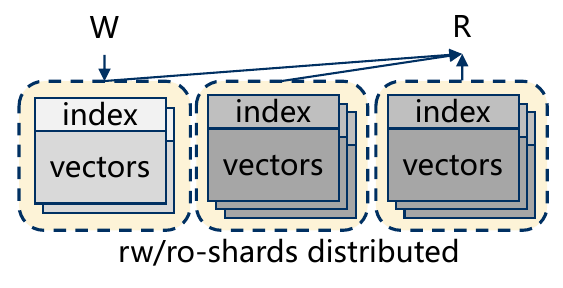}}
  \subfloat[\namesys{}\label{fig:typical-arch-d}]{\includegraphics[width=.5\textwidth]{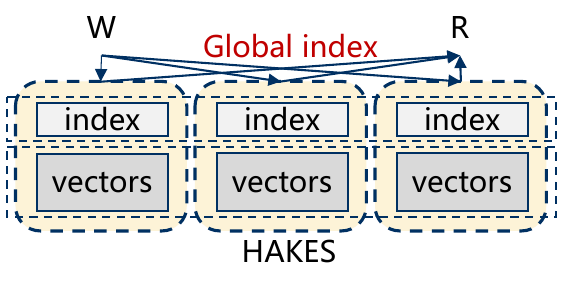}}
  \end{minipage}
  \hspace{-.4cm}
  \begin{minipage}[c]{0.1\textwidth}
  \centering
  \subfloat{\includegraphics[width=\textwidth]{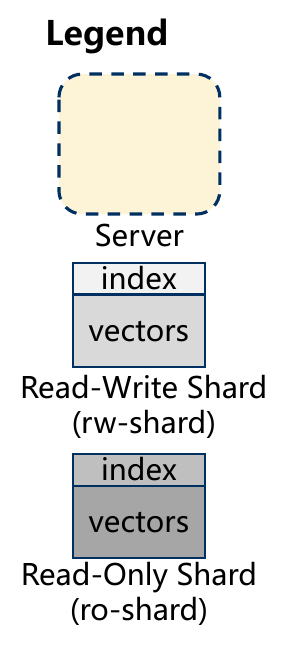}}
  \end{minipage}
  \vspace{-0.3cm}
  \hspace{-.3cm}
  \caption{Different architectures of distributed vector databases.} 
  \Description[short description]{long description}
  \label{fig:common-arch}
\end{figure}

\vspace{0.1cm}
\noindent{\bf Discussion.} \namesys{}'s architecture is different from that of existing distributed vector databases. Figure~\ref{fig:common-arch} compares four architectures with distinct shard layouts and communication for read and write. 
In the first architecture (Figure~\ref{fig:typical-arch-a}), adopted by~\cite{weaviate,lanns,qdrant}, each server hosts a single read-write shard and maintains its index. A read request merges search results from every node, while a write request is routed to a single server based on a sharding policy. In the second architecture (Figure~\ref{fig:typical-arch-b}), used by~\cite{cassandra-5.0}, each node maintains one read-write shard and multiple read-only shards to reduce the read-write contention. The third architecture in Figure~\ref{fig:typical-arch-c} extends the first two by employing multiple read-write shards and multiple read-only shards. 
It is adopted by~\cite{milvus,manu}, and supports scaling out of read or of write by adding servers for the required type of shards.
We note that in these three architectures, an index is local to the shard data, i.e., the index of each shard is not constructed over the global set of vectors. 
However, building many small indexes over multiple shards incurs significant overhead, as we show in our evaluation later.
\namesys{}'s architecture in Figure~\ref{fig:typical-arch-d}, in contrast, maintains
the global index at each server, since the filter stage index is small due to compression and supports efficient update.

The index in the filter stage scales with dataset size. 
However, \namesys{}'s high compression ratio enables a single cloud server to host TB-scale indexes.
For deployments where the index exceeds individual server capacity, the index is dynamically sharded across IndexWorker groups.
Searches query one replica per shard group while updates propagate atomically to all replicas in the affected group.
Full-precision vectors remain managed separately by RefineWorker nodes deployed on distinct servers, ensuring physical isolation between filter and refine stages.

\subsection{\namesys{} Design}

The \textit{IndexWorker} maintains a replica of the filter-stage index and the compressed vectors organized in IVF partitions. It takes a query vector as input and returns a set of candidate vectors. IndexWorker is compute-heavy. It implements dynamic batching with internal, lock-free task queues. In particular,  vectors from different requests are batched into a matrix such that the dimensionality reduction and IVF assignment can be computed efficiently via matrix-matrix multiplication. Requests are batched only under high load, otherwise, they are processed immediately on separate CPU cores. 

The \textit{RefineWorker} maintains a shard of the original vectors. It handles the refine stage, which evaluates similarity scores between the query and the candidate vectors belonging to the shard. \namesys{} supports two types of sharding policies for the full vectors. One policy is sharding by vector ID, in which vectors are distributed (evenly) among the nodes by their IDs. The other is sharding by IVF assignment, in which vectors belonging to the same IVF partition are on the same RefineWorker. 
This policy helps reduce network communication
because the refine stages only happen on a small number of nodes. 

\vspace{0.1cm}
\noindent{\bf Operation workflow.}
Before serving queries, \namesys{} builds an index over a given dataset. It first takes a representative sample of the dataset to initialize the base index parameters. It then launches IndexWorkers that use the base index. Next, it inserts the vectors, and after that starts serving search requests. 
It builds training datasets for learning index parameters by collecting the results of ANN queries. Once the training process finishes, it installs the new parameters to all IndexWorkers with minimal disruption. Specifically, at every IndexWorker node, the new parameters are loaded to memory and the pointers in \nameindex{} are redirected to them.

During search, the client sends the query to an IndexWorker and gets back the candidate vectors. Based on the sharding configuration, the client sends these vectors to the corresponding RefineWorkers in parallel. 
The client reranks the vectors returned by the RefineWorkers and outputs the top $k$ vectors. 
During insert, the client sends the new vector to the RefineWorker that manages the shard where the vector is to be inserted. The client then picks an IndexWorker to compute the new quantized vector and update the IVF structure. 
This update is broadcast to all the IndexWorkers.
For deletion, the client broadcasts the vector IDs to be deleted to all the IndexWorkers, which then mark them as deleted in their filter-stage index.

\vspace{0.1cm}
\noindent{\bf Consistency and failure recovery}
\namesys{} does not guarantee strong consistency, which is acceptable because applications relying on vector search can tolerate that
~\cite{milvus, analyticdb-v}. 
It can support session consistency by synchronously replicating the write requests or having the client stick to an IndexWorker.
\namesys{} periodically creates checkpoints of the index. 
During crash recovery, new vectors after the checkpoints are re-inserted into the RefineWorkers and IndexWorkers.

\section{Evaluation}
\label{sec:evaluation}

In this section, we 
benchmark \nameindex{} against state-of-the-art ANN indexes and \namesys{} against state-of-the-art distributed vector databases to study the effectiveness of our design

\subsection{Implementation}
\label{subsec:eval:impl}
We implement \nameindex{} by extending the FAISS library~\cite{faiss}.
IndexWorker and RefineWorker are implemented on top of the index \nameindex{}, and they are accessible via an HTTP server implemented using libuv and llhttp.
The index extension and serving system take $\sim7000$ LoC in C++. The index training is implemented in $\sim1000$ LoC in Python, using Pytorch@1.12.1. The \namesys{} client is implemented in Python in $\sim500$ LoC.

\subsection{Experiment Setup}
\label{subsec:eval:setup}
\noindent\textbf{Datasets and workloads.}
As listed in Table~\ref{tab:1m-datasets}, we use six deep embedding datasets and the GIST dataset.
Five of the datasets are at the 1-million scale, and we use them for index benchmarking.
\begin{itemize}[leftmargin=*]
    \item \underline{DPR-768} is generated by the 
    Dense Passage Retrieval (DPR) context encoder model~\cite{dpr}  on text records sampled from the Sphere web corpus dataset~\footnote{https://weaviate.io/blog/sphere-dataset-in-weaviate}.
    \item \underline{OPENAI-1536}~\cite{dbpedia-openai-data} is generated by OpenAI's embedding service on DBpedia text data~\cite{beir}. 
    
    \item \underline{MBNET-1024}: is generated by pretrained MobileNet~\cite{mobilenet} on one million ImageNet data~\cite{imagenet}.
    \item \underline{RSNET-2048}: is generated by pretrained ResNet~\cite{resnet} on 1 million ImageNet data.
    \item \underline{GIST-960}: is a widely in the literature for benchmarking ANN indexes~\cite{bench-tkde19, ann-benchmarks, adsampling}. We selected GIST for its high dimensionality. 
\end{itemize}

\noindent We also use two other large datasets for in-depth analysis of our index and system. 
\begin{itemize}[leftmargin=*]
    \item \underline{DPR-768-10m}: use the same embedding model as DPR-768 but on 10 million Sphere text records.
    \item \underline{E5-1024-10m}: 
    is generated with the E5-large text  model~\cite{e5} on 10 million Sphere text records.
\end{itemize}

\noindent We normalize the vectors and use the inner product as the similarity metric due to its popularity in existing embedding services
\footnote{https://platform.openai.com/docs/guides/embeddings} 
\footnote{https://docs.mistral.ai/capabilities/embeddings/}.
This metric is also the default choice in all of the baseline systems~\cite{milvus, weaviate, cassandra-5.0}.
We note that for normalized vectors, Euclidean distance, cosine similarity, and inner product are equivalent with respect to neighbor relationships. 
The search quality is measured by Recall10@10.
The ground-truth nearest neighbors for the queries are generated by a brute-force search over the entire dataset.

\vspace{0.1cm}
\noindent{\textbf{Training setup.}} Index training is conducted on an Ubuntu 18.04 server that has an Intel Xeon W-2133@3.60GHz CPU with 6 cores and an NVIDIA GeForce RTX 2080 Ti GPU. The $\lambda$ parameter is searched in the set $\{0.01, 0.03, ... 30\}$. The AdamW Optimizer is used with a learning rate value in the set $\{10^{-5}, 10^{-4}, 10^{-3}\}$. The batch size is set to 512. We use 100,000 samples and their 50 neighbors returned by the base index at $nprobe=1/10$ and $k'/k = 10$.

\vspace{0.1cm}
\noindent{\textbf{Environment setup.}} 
We conduct all the index experiments on a Ubuntu 20.04 server equipped with an Intel Xeon W-1290P @ 3.70GHz CPU with 10 cores and 128 GiB memory.
We run distributed experiments in a cluster of servers with the above hardware specification.

\begin{table}[t]
    \centering
    \caption{High-dimensional datasets.}
    \vspace{-0.3cm}
    \begin{tabular}{|c|c|c|c|c|c|}
    \hline
    Dataset & N & d & nq & Size (GiB) & Type \\\hline\hline
    DPR-768 & 1000000 & 768 & $10^4$ & 2.86 & Text \\\hline
    OPENAI-1536 & 990000 & 1536 & $10^4$ & 5.72 & Text \\\hline
    MBNET-1024 & 1103593 & 1024 & $10^3$ & 4.21 & Image \\\hline
    RSNET-1024 & 1103593 & 2048 & $10^3$ & 8.42 & Image \\\hline
    GIST-960 & 1000000 & 960 & $10^3$ & 3.58 & Image \\\hline
    DPR-768-10m & 9000000 & 768 & $10^6$ & 26 & Text \\\hline
    E5-1024-10m & 9000000 & 1024 & $10^6$ & 35 & Text \\
    \hline
    \end{tabular}
    \label{tab:1m-datasets}
\end{table}

\vspace{0.1cm}
\noindent{\textbf{Index  baselines.}}
We select 12 state-of-the-art in-memory ANN index baselines, covering both IVF partitioning-based and graph-based indexes. 
\begin{itemize}[leftmargin=*]
\item \underline{IVF} is the classic IVF index with k-means clustering.
\item \underline{IVFPQ\_RF} applies PQ with IVF and uses FAISS 4-bit quantization fast scan implementation~\cite{fastscan, faiss}. The RF denotes a refine stage
\item \underline{OPQIVFPQ\_RF} uses OPQ~\cite{opq} to learn a rotation matrix
that minimizes the PQ reconstruction error. {We use the OPQ implementation in FAISS that can generate a transformation matrix $\in \mathbb{R}^{d\times d_r}$ to reduce dimension before IVF and PQ.}
\item \underline{HNSW} ~\cite{hnsw} is the index used in almost all vector databases. 
\item \underline{ELPIS}~\cite{elpis} partitions the dataset and maintains an HNSW graph index for each partition, representative for maintaining multiple subgraph indexes.
\item \underline{LSH-APG}~\cite{lsh-apg} leverages LSH to identify close entry points on its graph index to reduce the search path length.
\item \underline{ScaNN}~\cite{scann}, \underline{SOAR}~\cite{soar}, and \underline{RaBitQ}~\cite{rabitq} are recent proposed quantization scheme used with partitioning-based index. {They use reranking to improve recall.}
\item \underline{Falconn++}~\cite{falconnlsf} and \underline{LCCS}~\cite{lccs-lsh} are state-of-the-art LSH indexes.
\item \underline{LVQ}~\cite{lvq} is a state-of-the-art graph index with quantization. {It supports 4-bit scalar quantization, followed by 8-bit quantization on residual for reranking.}
\end{itemize}
\noindent We do not compare against recent indexes that are optimized for secondary storage~\cite{SPANN, starling}, which report lower performance than in-memory indexes.
{The first three IVF baselines share the codebase of our extended FAISS library. For the remaining baselines, we use the implementations provided by the authors.}

\vspace{0.1cm}
\noindent{\textbf{Distributed vector database baselines.}} We select three popular distributed vector databases that employ in-memory ANN indexes. 
They cover the three architectures described in Section~\ref{subsec:sys:arch}. 
We set up the systems according to the recommendations from their respective official documentation.
\begin{itemize}[leftmargin=*]
    \item \underline{Weaviate~\cite{weaviate}} adopts an architecture in which each server maintains a single read-write shard and an HNSW graph. It implements HNSW natively in Golang with fine-grained node-level locking for concurrency. We deploy Weaviate using the official Docker image at version v1.21.2

    \item \underline{Cassandra} adds support for vector search recently~\cite{cassandra-5.0} on its NoSQL database. It shards the data across nodes, and every node maintains a read-write shard and multiple read-only shards that are periodically merged. It uses jVector~\footnote{https://github.com/datastax/jvector}, a graph index that only searches quantized vectors, similar to DiskANN~\cite{diskann}.
    
    \item \underline{Milvus~\cite{milvus, manu}} adopts an architecture in which there is one shard that processes updates. Once reaching 1GiB, this shard becomes a read-only shard with its own index and is distributed across the servers for serving. We deploy Milvus version 2.4 using the official milvus-operator v0.9.7 on a Kubernetes (v1.23.17) cluster.
\end{itemize}
Besides the three systems above, we add two more baselines, called \underline{Sharded-HNSW} and \underline{HAKES-Base}. Sharded-HNSW adopts Weaviate's architecture, and uses our server implementation with hnswlib. This baseline helps isolate the performance impact of the index and system design, since the three vector databases are implemented with different languages and have different sets of features.  HAKES-Base is the same as \namesys{} but employs the base index, that is, without parameter training or optimizations.

\begin{figure*}[t]
  \centering
  \includegraphics[width=0.98\textwidth]{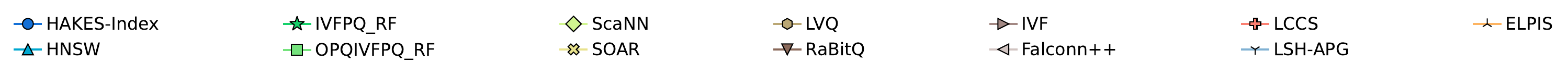}\vspace{-5mm}\\
    
  \subfloat[{DPR-768} \label{fig:sphere-768-1m-recall80}]{\includegraphics[width=0.195\textwidth]{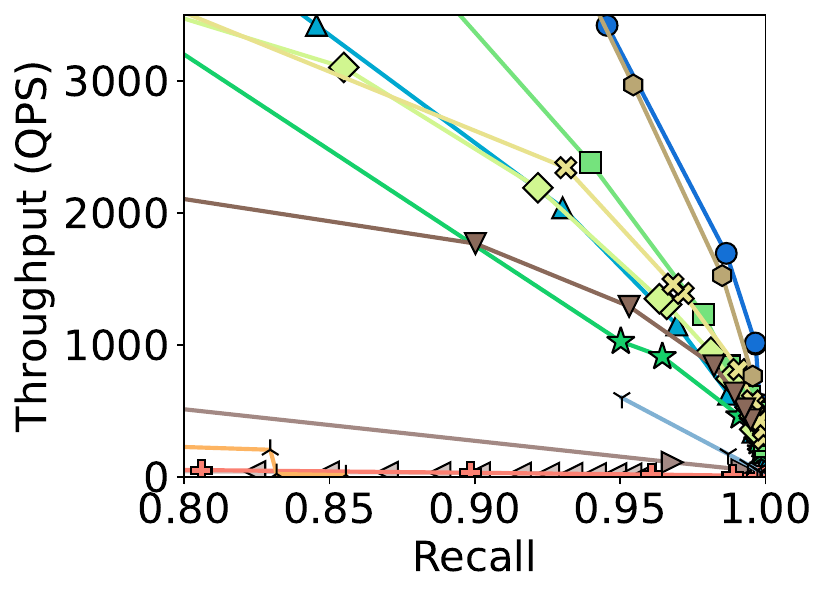}}
  \subfloat[{OPENAI-1536} \label{fig:DBpedia-openai-recall80}]{\includegraphics[width=0.195\textwidth]{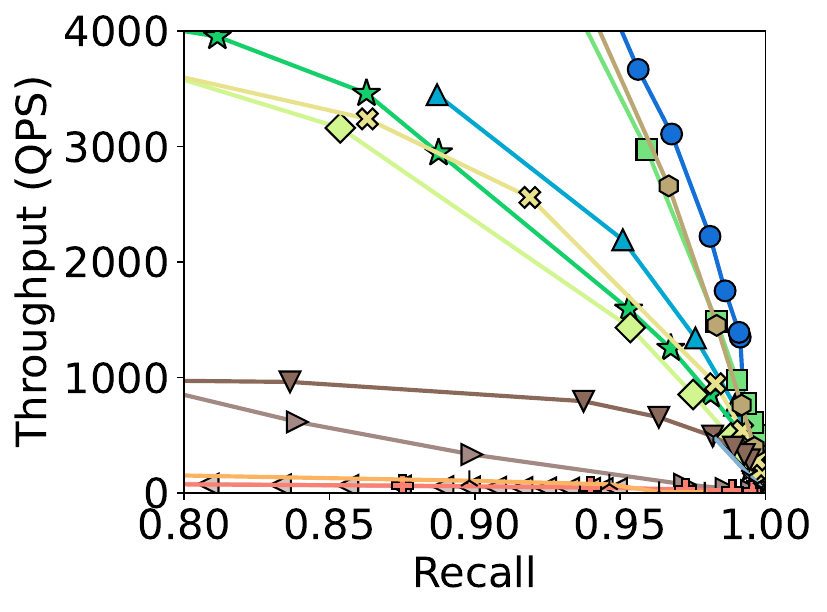}}
  \subfloat[{MBNET-1024} \label{fig:mbnet-1024-recall80}]{\includegraphics[width=0.195\textwidth]{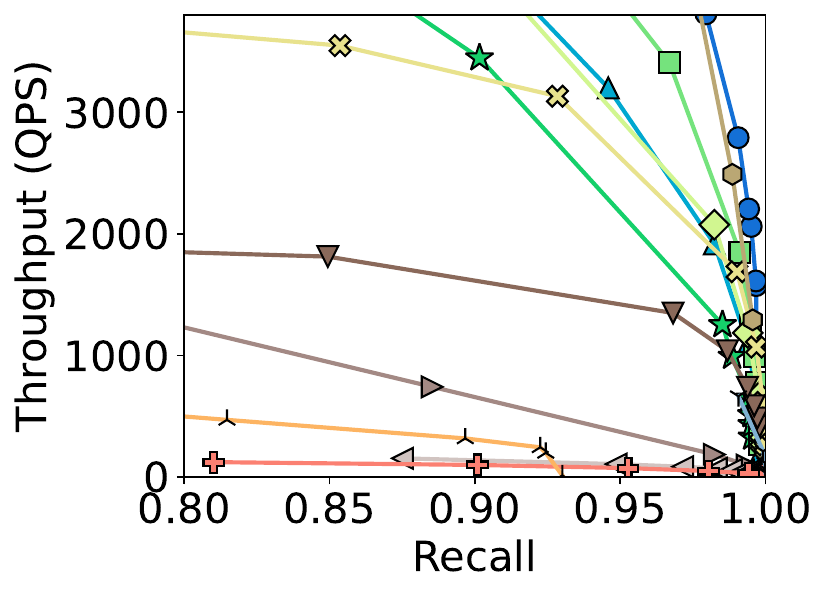}}
  \subfloat[{RSNET-2048} \label{fig:rsnet-2048-recall80}]{\includegraphics[width=0.195\textwidth]{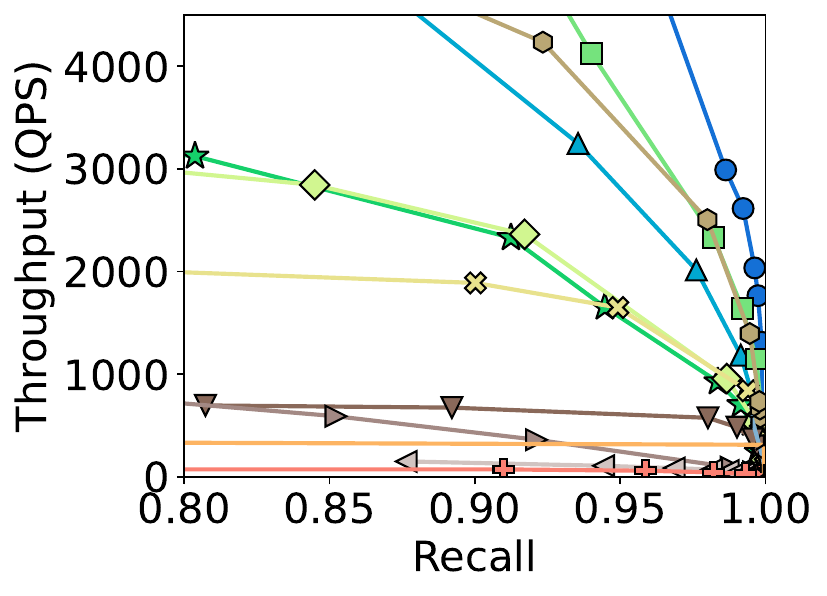}}
  \subfloat[{GIST-960} \label{fig:gist-960-recall80}]{\includegraphics[width=0.195\textwidth]{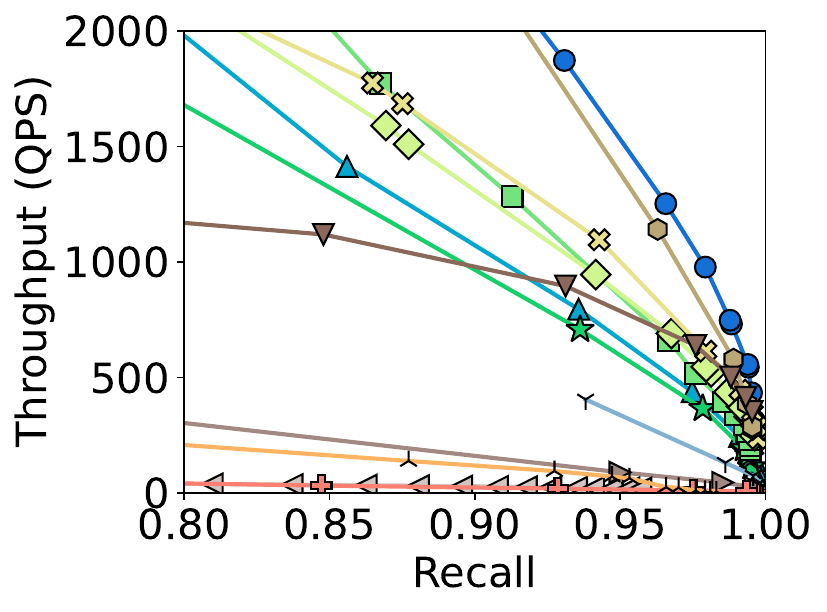}}
  \vspace{-0.3cm}
  \caption{{Throughput vs. recall for sequential reads (recall $\ge$ 0.8).}}
  \Description[short description]{long description}
  \vspace{-2mm}
  \label{fig:index-recall}
\end{figure*}

\begin{figure*}[t]
  \centering
  \includegraphics[width=0.98\textwidth]{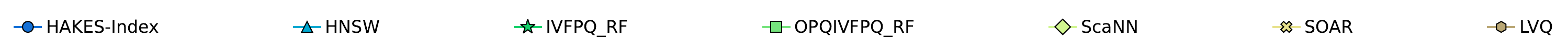}\vspace{-5mm}\\
  
  \subfloat[DPR-768\label{fig:dpr-768-seq-rw}]{\includegraphics[width=0.195\textwidth]{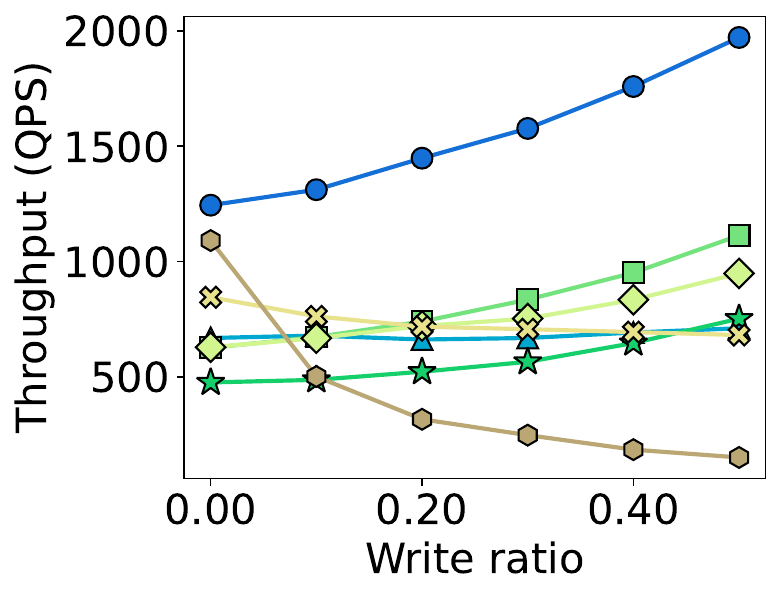}}
  \subfloat[OPENAI-1536\label{fig:openai-1536-seq-rw}]{\includegraphics[width=0.195\textwidth]{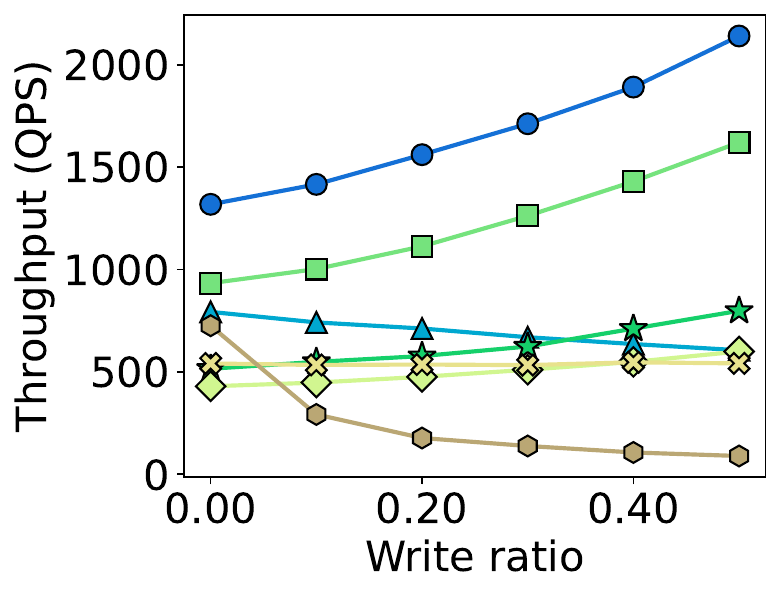}}
  \subfloat[MBNET-1024\label{fig:mbnet-1024-seq-rw}]{\includegraphics[width=0.195\textwidth]{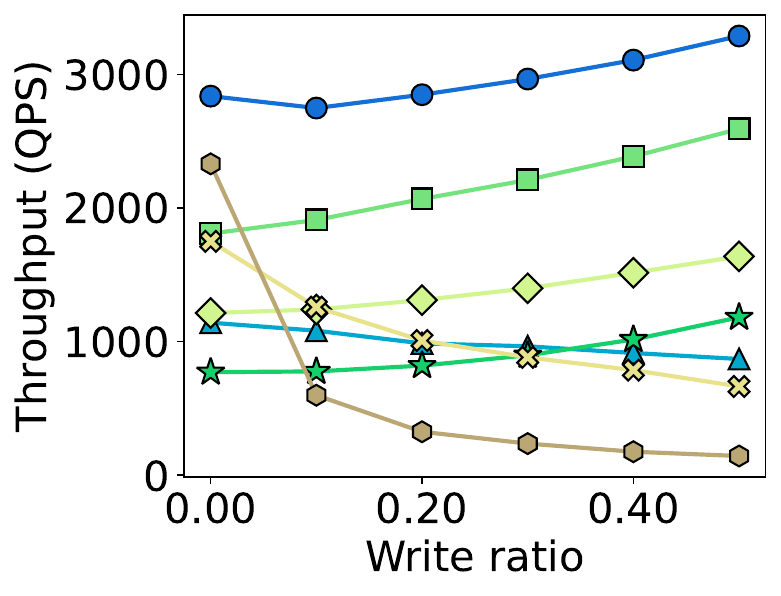}}
  \subfloat[RSNET-2048\label{fig:rsnet-2048-seq-rw}]{\includegraphics[width=0.195\textwidth]{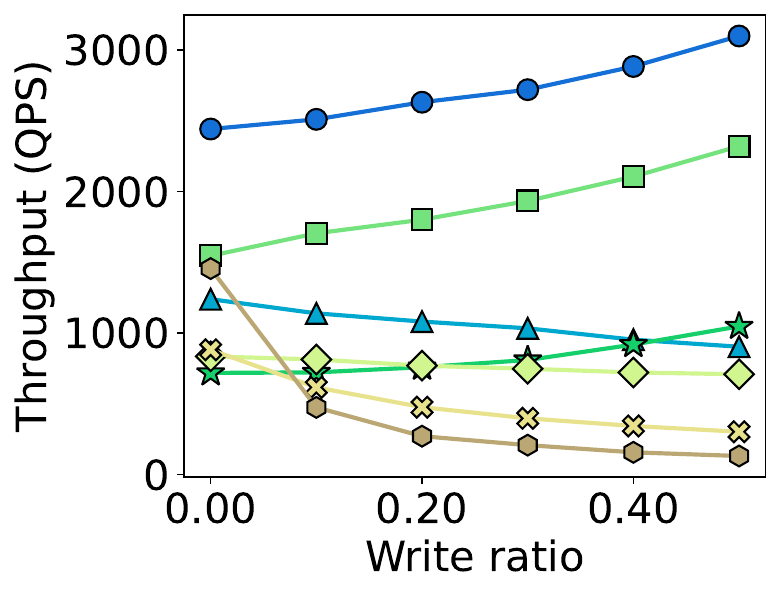}}
  \subfloat[GIST-960\label{fig:gist-960-seq-rw}]{\includegraphics[width=0.195\textwidth]{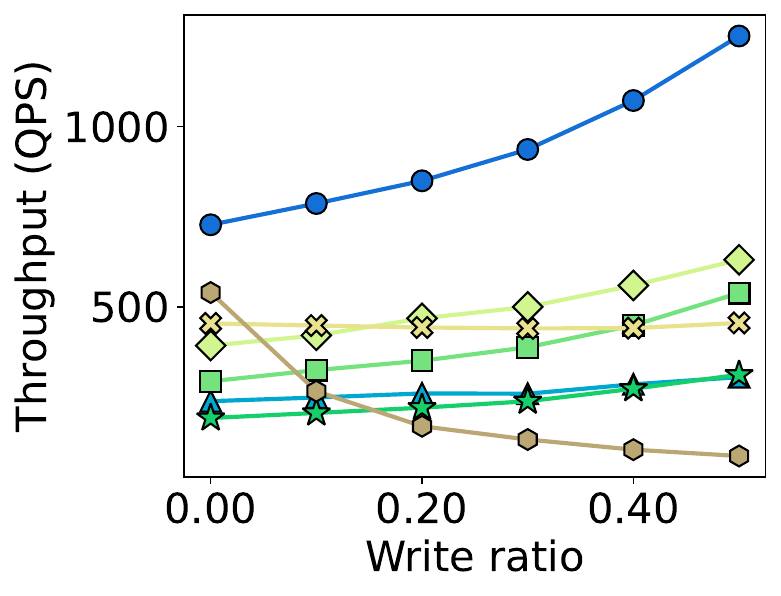}}
  \vspace{-0.3cm}
  \caption{Performance under sequential read-write workloads. (Recall=0.99).}
  \Description[short description]{long description}
  \label{fig:index-seq-rw}
\end{figure*}

\begin{figure*}[t]
  \centering
  \vspace{-0.2cm}
  \includegraphics[width=0.60\textwidth]{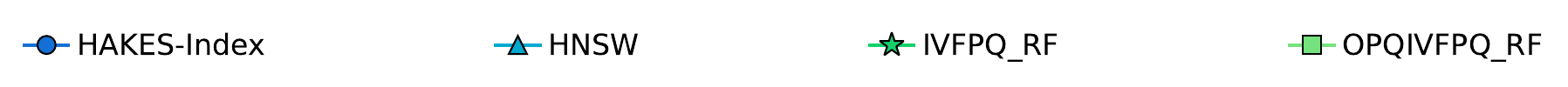}\vspace{-5mm}\\
  
  \subfloat[DPR-768\label{fig:dpr-768-rw}]{\includegraphics[width=0.195\textwidth]{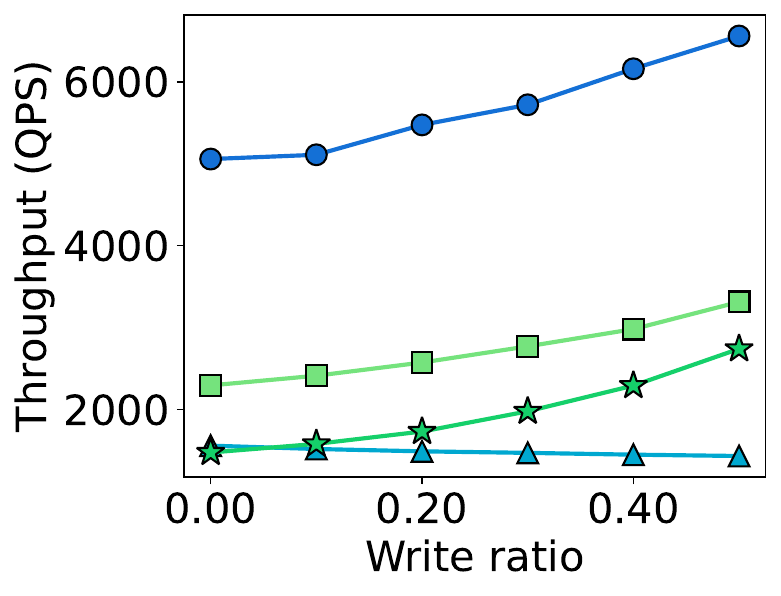}}
  \subfloat[OPENAI-1536\label{fig:openai-1536-rw}]{\includegraphics[width=0.195\textwidth]{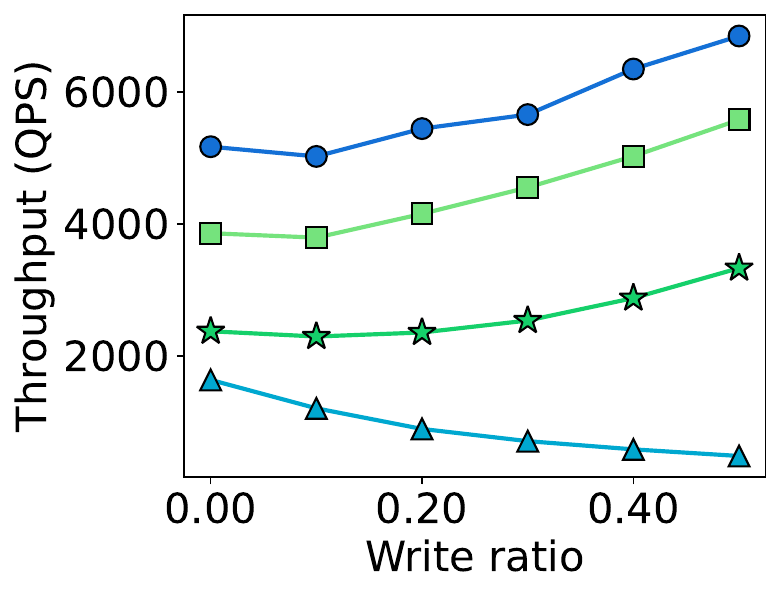}}
  \subfloat[MBNET-1024\label{fig:mbnet-1024-rw}]{\includegraphics[width=0.195\textwidth]{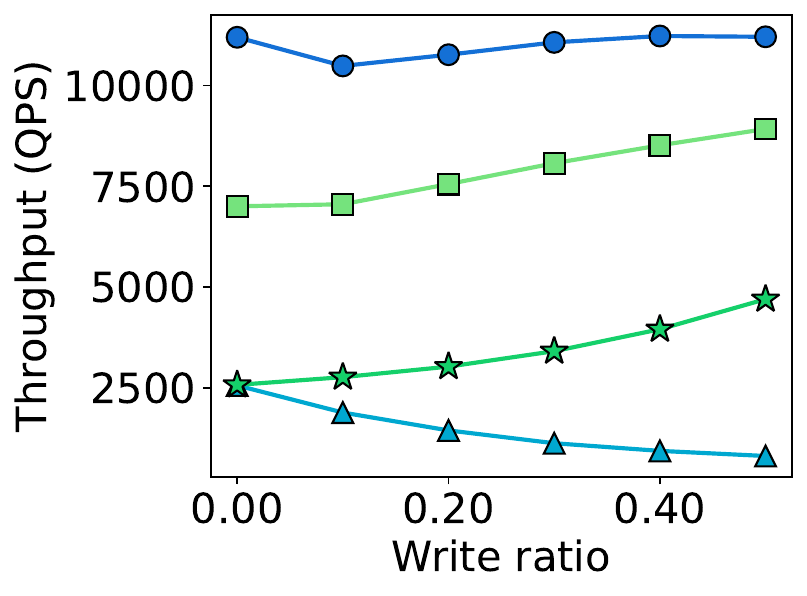}}
  \subfloat[RSNET-2048\label{fig:rsnet-2048-rw}]{\includegraphics[width=0.195\textwidth]{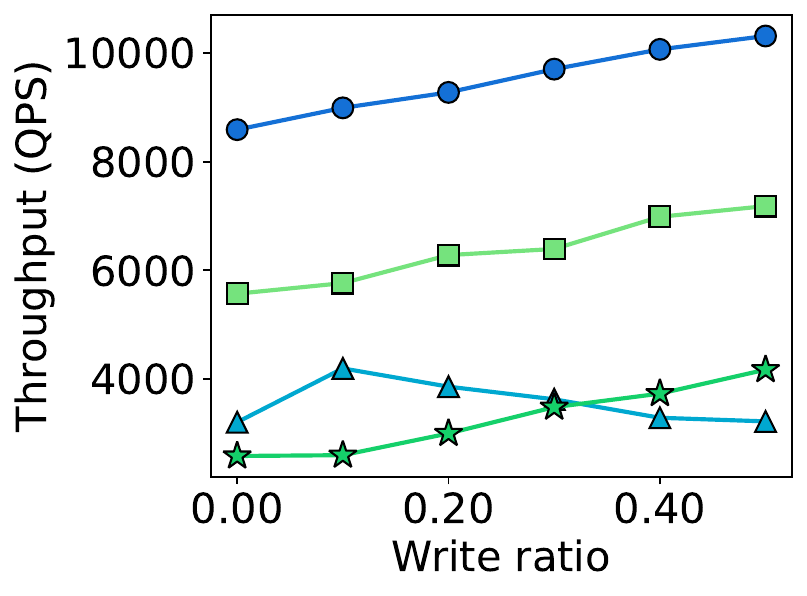}}
  \subfloat[GIST-960\label{fig:gist-960-rw}]{\includegraphics[width=0.195\textwidth]{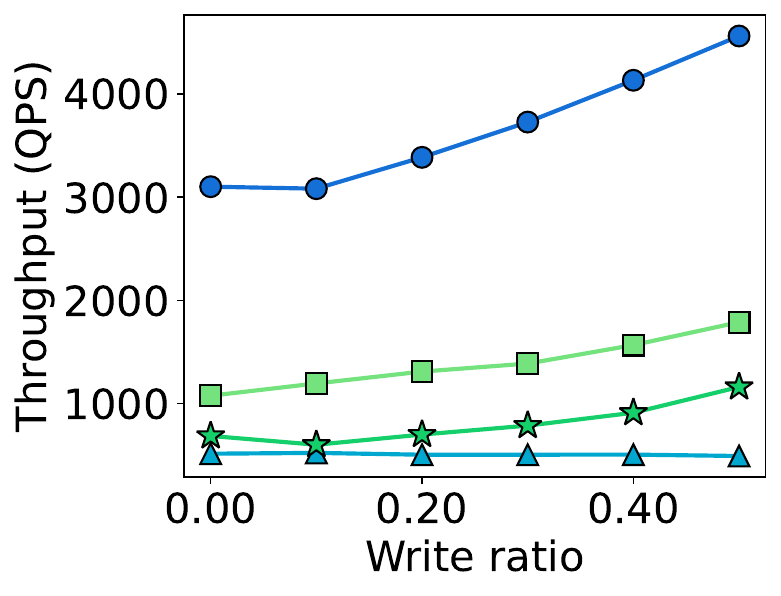}}
  \vspace{-0.3cm}
  \caption{Performance under concurrent read-write workloads. (Recall=0.99).}
  \vspace{-0.2cm}
  \Description[short description]{long description}
  \label{fig:index-con-rw}
\end{figure*}

\subsection{Index Benchmarking and Analysis}
\label{subsec:eval:bench}

{
For each index, we explore the range of configurations recommended in the original paper and corresponding code repository,}
and pick the best configuration for each dataset.
We then run experiments with varying search parameter values to examine the index's throughput-recall tradeoff.
The complete set of explored and selected configurations of all indexes is in the Appendix.

\vspace{0.1cm}
\noindent\textbf{Sequential read workload.}
Figure~\ref{fig:index-recall} compares the throughput-recall tradeoff of the $13$ indexes for the recall range above $80\%$.
Across the different datasets, \nameindex{}{} achieves state-of-the-art throughput-recall tradeoff.  At high recall, it even outperforms the recent quantized graph index, LVQ, which is heavily optimized for prefetching and SIMD acceleration.
The performance difference among OPQIVFPQ\_RF, IVFPQ\_RF, and IVF confirms that with a refine stage, deep embeddings can be compressed significantly with quantization and dimensionality reduction for efficiency while maintaining high accuracy.
Across the deep embedding datasets, OPQIVFPQ\_RF and \nameindex{} achieve the reported tradeoff with $d_r/d = $1/4 or 1/8, significantly reducing computation.

Recent quantization-based indexes, namely ScaNN, SOAR, and RaBitQ, show mixed results compared to IVFPQ\_RF, which uses standard PQ and fast scan implementation. ScaNN improves the quantization for inner product approximation; SOAR aims to reduce the correlation of multiple IVF partitions assignment for one vector; and RaBitQ uses LSH to generate binary code representation and {
decide vectors to be reranked with its error bound.} ScaNN and SOAR outperform IVFPQ\_RF on GIST-960, DPR-768, and MBNET-1024, but have comparable performance on RSNET-2048 and OPENAI-1536. 
RaBitQ only performs better than IVFPQ\_RF on GIST-960. These observations highlight the importance of evaluating indexes on high-dimensional deep embedding. 

The performance of Falconn++ and LCCS ranks below IVF, confirming that LSH-based indexes are less effective in filtering vectors than the data-dependent approaches in high-dimensional space~\cite{ann-benchmarks, bench-tkde19, rabitq, lsh-apg}.
Among graph-based indexes, {LVQ performs best as its scalar quantization avoids computation using full vectors for graph traversal.} 
The difference between HNSW and LSH-APG indicates that the hierarchical structure of HNSW is more effective than the LSH-based entry point selection in LSH-APG in high-dimensional space. The gap between HNSW and ELPIS shows that sharding a global graph index into smaller subgraphs degrades the overall performance. We analyze that phenomenon in distributed vector databases in Section~\ref{subsec:eval:sys}.

\vspace{0.1cm}
\noindent\textbf{Read-write workload.}
For indexes supporting inserts, we first evaluate their performance with sequential read-write workloads. We focus on high-recall regions of $~0.99$, and vary the write ratio from $0.0$ to $0.5$. Figure~\ref{fig:index-seq-rw} reveals that as the write ratio increases, partitioning-based indexes have a clear advantage over graph indexes. Both LVQ's and HNSW's performance decrease as the write ratio increases, because inserting new data into a  graph is slower than serving an ANN search. The reverse is true for partitioning-based indexes, since insert does not involve comparison with existing vectors. The exceptions are ScaNN in RSNET and SOAR, which select quantized code with additional constraints. In particular, SOAR assign a vector to multiple partitions based on their correlation which is more costly than a single assignment used by other partitioning-based indexes.
\nameindex{} outperforms all baselines across all datasets, because of its efficient search and insert. 

We further evaluate the indexes supporting  concurrent read-write workloads. The HNSW implementation in hnswlib supports concurrent read-write with fine-grained locking on the graph nodes, and using our extension on FAISS, IVFPQ\_RF, and OPQIVFPQ\_RF also support partition locking as our index does.
We use 32 clients and vary the ratio of write requests.
Figure~\ref{fig:index-con-rw} shows that partitioning-based indexes are better than HNSW, due to low contention and predictable memory access pattern. We note that even IVFPQ\_RF reaches a comparable or higher throughput than HNSW for concurrent read. The performance gaps increase with more writes.

\vspace{0.1cm}
\noindent\textbf{Memory consumption.}
We observe that the cost of storing the original vectors dominates the index's memory consumption. We discuss the memory overhead for representative baselines on OPENAI-1536 as an example. {We measure memory usage before and after loading the indexes}. 
HNSW maintains {the connection information for each node at each level} on top of the original data, increasing the memory from 5.72 to 6.01 GiB.
For IVFPQ\_RF, OPQIVFPQ\_RF, and \nameindex{}, the main overhead is storing the compressed vectors. IVFPQ\_RF consumes 5.92 GiB, where OPQIVFPQ\_RF consumes 5.86 GiB due to dimensionality reduction.  
\nameindex{} consumes 5.86 GiB similar to OPQIVFPQ\_RF, 
as the additional set query index parameters are small.

\subsection{\nameindex{} Analysis}
\label{subsec:eval:index}

\begin{table}[]
    \centering
    \caption{{Ablation study where recall is in the 0.99 region. Each cell shows the QPS (recall) value.}}
    \label{tab:app:abs-so}
    \vspace{-0.3cm}
    \resizebox{0.475\textwidth}{!}{
    
    \begin{tabular}{|c|c|c|c|c|}
        \hline
          & Base & Learn & Learn + SQ & All  \\ \hline
        {DPR-768} 
        & 1233 (0.979) & 1238 (0.991) & 1280 (0.990) & 1280 (0.990) \\ \hline
        {OPENAI-1536} 
        &  977 (0.990) & 974 (0.994) & 976 (0.994) & 1389 (0.991) \\ \hline
        {MBNET-1024} 
        &  2393 (0.977) & 2413 (0.991) & 2579 (0.991) & 2791 (0.991) \\ \hline
        {RSNET-1024} 
        &  2330 (0.982) & 2350 (0.992) & 2444 (0.992) & 2615 (0.992) \\ \hline
        {GIST-960} 
        &  610 (0.944) & 625 (0.989) & 631 (0.989) & 747 (0.988) \\ \hline
    \end{tabular}
    }
\end{table}

\vspace{0.1cm}
\noindent\textbf{Performance gain breakdown.}
Table~\ref{tab:app:abs-so} shows how the different techniques contribute to the performance of \nameindex{}. We report the results at the search configurations that achieve $recall \approx 0.99$ with the learned parameters. The learned compression contributes the most as it improves the throughput-recall tradeoff over the base settings. Scalar quantization of IVF centroids and early termination provide further improvement to throughput without significantly degrading recall. We used the same setting $t=k'/200$ and $n_t=30$ for early termination, which improves the throughputs considerably on 4 of the 5 datasets. However, as discussed in the previous subsection, the heuristic can terminate the search prematurely and miss the true close neighbors, thereby leading to lower recall.
We note that careful tuning on a dataset can achieve better performance for a specific recall target. 

\begin{table*}[t]
    \centering
    \caption{Recall improvement at different search configurations.}
    \label{tab:index-enhance}
    \vspace{-0.2cm}
\resizebox{\textwidth}{!}{
    \begin{tabular}{|c|c|ccc|ccc|ccc|ccc|}
    \hline
    \multicolumn{2}{|c|}{IVF $nprobe$ (total 8192)}  & \multicolumn{3}{c|}{200} & \multicolumn{3}{c|}{400} & \multicolumn{3}{c|}{600} & \multicolumn{3}{c|}{800}\\
    \hline
    \multicolumn{2}{|c|}{$k'/k$} & 10 & 50 & 200 & 10 & 50 & 200 & 10 & 50 & 200 & 10 & 50 & 200 \\ \hline
    \multirow{2}{*}{DPR-768-10m} 
        & Base & 0.722 & 0.904 & 0.968 & 0.725 & 0.909 & 0.976 & 0.725 & 0.910 & 0.979 & 0.725 & 0.911 & 0.980 \\ 
        & Learned    & \textbf{0.859} & \textbf{0.963} & \textbf{0.980} & \textbf{0.866} & \textbf{0.973} & \textbf{0.993} & \textbf{0.868} & \textbf{0.976} & \textbf{0.996} & \textbf{0.869} & \textbf{0.976} & \textbf{0.997} \\ \hline
    \multirow{2}{*}{E5-1024-10m} 
        & Base & 0.765 & 0.896 & 0.942 & 0.773 & 0.910 & 0.959 & 0.777 & 0.914& 0.966 & 0.778 & 0.917 & 0.969 \\ 
        & Learned & \textbf{0.843} & \textbf{0.932} & \textbf{0.953} & \textbf{0.856} & \textbf{0.950} & \textbf{0.974} & \textbf{0.860} & \textbf{0.955} & \textbf{0.979} & \textbf{0.862} & \textbf{0.958} & \textbf{0.983}\\ \hline
    \end{tabular}
    }
\end{table*}

\vspace{0.1cm}
\noindent\textbf{Recall improvement by learned compression.}
Table~\ref{tab:index-enhance} reports the recalls for different search configurations, for 
10-million scale datasets. We note that the training process does not affect the cost of performing dimensionality reduction and of scanning quantized vectors. In other words, given the same search parameters, the performance is only affected by the IVF partition selection, which we observe to be negligible. 
Table~\ref{tab:index-enhance} shows consistent improvement across all configurations on the 10-million scale datasets. The improvement is between 0.07 to 0.14 for $k'/k=10$ and 0.01 to 0.07 for those settings with recall over 0.9. It is higher for smaller filter candidate sets (i.e. smaller $k'/k$), which is expected because the impact of high-quality candidate vectors is higher when the candidate set is small. This improvement allows \nameindex{} to reach high recall with a smaller $nprobe$ and $k'/k$, which translates to higher throughput. We attribute the high recalls to the training process that results in the refine stage having more true nearest neighbors.
We discuss the results on a 1-million scale dataset in the Appendix.

\vspace{0.1cm}
\noindent\textbf{Training cost.} 
The cost of constructing \nameindex{} consists of the cost of building the base index, and of training the compression parameters. Deploying the trained parameters incurs negligible overhead, as it only loads a small dimensionality reduction matrix, bias vector, quantization codebooks, and IVF centroids into memory. For the 10-million scale datasets, building the base index takes 179.2s and 219.22s for initializing the OPQ and IVF parameters, and 103.2s and 125.6s to insert the 10 million vectors for DPR-768-10m and E5-1024-10m, respectively. 
It takes 52.9s and 60.9s for the two datasets respectively to sample the training set with 1/100 ratio and compute the approximate nearest neighbors with $nprobe$ set to the 1/10 partitions and $k'/k$. Training takes 34.9s and 45.6s, respectively. When deployed on a cluster, the time to insert vectors and prepare the training set neighbors can be reduced linearly with the number of nodes. 
In comparison, constructing the HNSW graph takes 5736.4s and 9713.21s on DPR-768-10m and E5-1024-10m datasets, which are $15.5\times$ and $21.5\times$ higher than the cost of building \nameindex{}. We note that in production, \nameindex{} can use the initialized parameters to serve requests, while training is conducted in the background using GPUs. The learned parameters can be seamlessly integrated once available, without rebuilding the index.

\begin{figure*}[ht]
  \centering
  \begin{minipage}[c]{0.42\textwidth}
    \includegraphics[width=\textwidth]{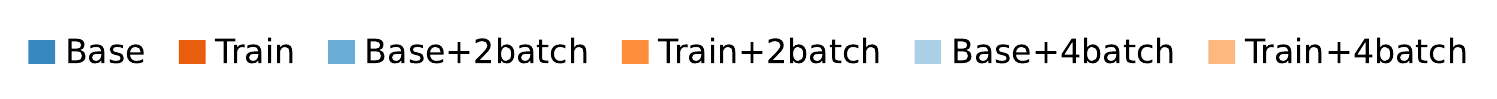}
    \vspace{-0.8cm}
    \\
    \subfloat[Recall\label{fig:drift-recall}]{\includegraphics[width=0.5\textwidth]{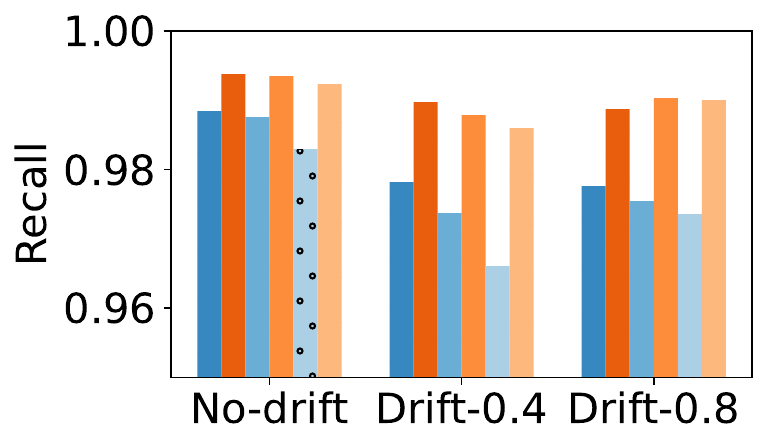}}
    \subfloat[Throughput\label{fig:drift-throughput}]{\includegraphics[width=0.5\textwidth]{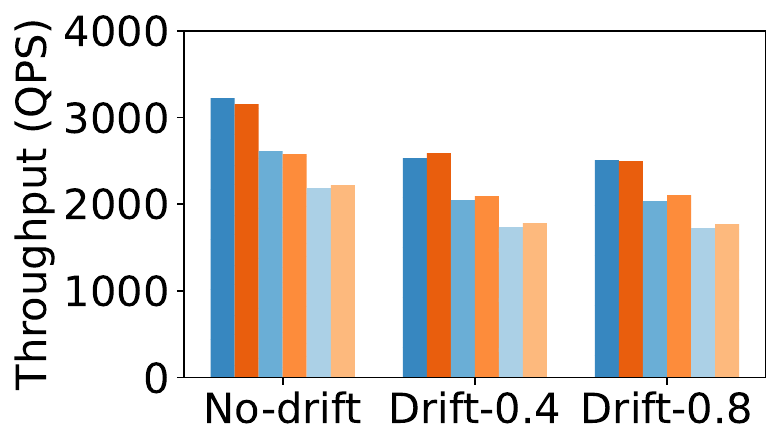}}
    \vspace{-0.2cm}
    \caption{Tolerance against data drift (MBNET-1024).}
  \Description[short description]{long description}
  \label{fig:ablation-drift}
  \end{minipage}
  \begin{minipage}[c]{0.33\textwidth}
  \centering
  \includegraphics[width=0.5\textwidth]{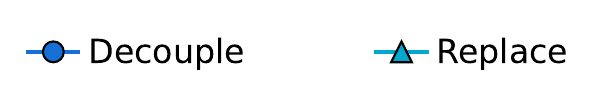}
  \vspace{-0.4cm}
  \\
  \subfloat[MBNET-1024\label{fig:mbnet-decouple-need}]{\includegraphics[width=0.5\textwidth]{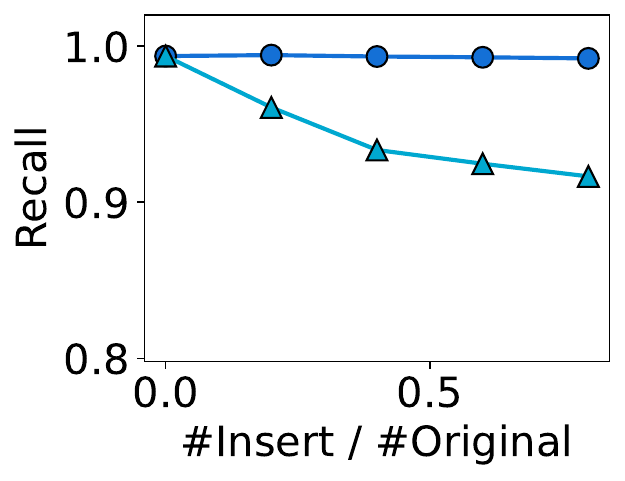}}
  \subfloat[RSNET-2048\label{fig:rsnet-decouple-need}]{\includegraphics[width=0.5\textwidth]{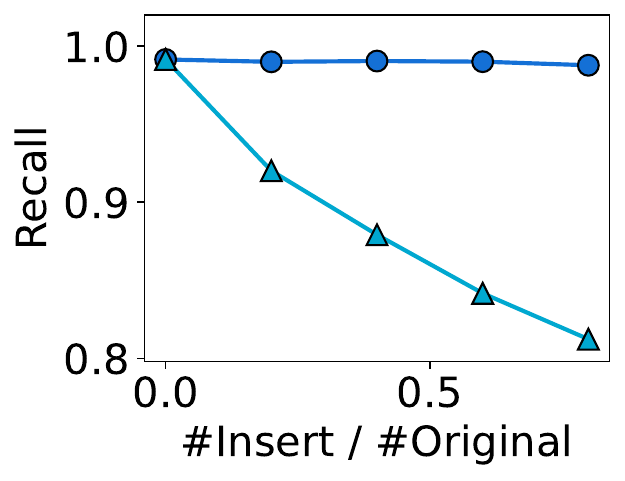}}
  \vspace{-0.2cm}
  \caption{Decoupling index parameters.}
  \Description[short description]{long description}
  \label{fig:ablation-decouple}
  \end{minipage}
  \begin{minipage}[c]{0.24\textwidth}
  \centering
    \subfloat[{R+D}\label{fig:all-delete}]{\includegraphics[width=0.5\textwidth]{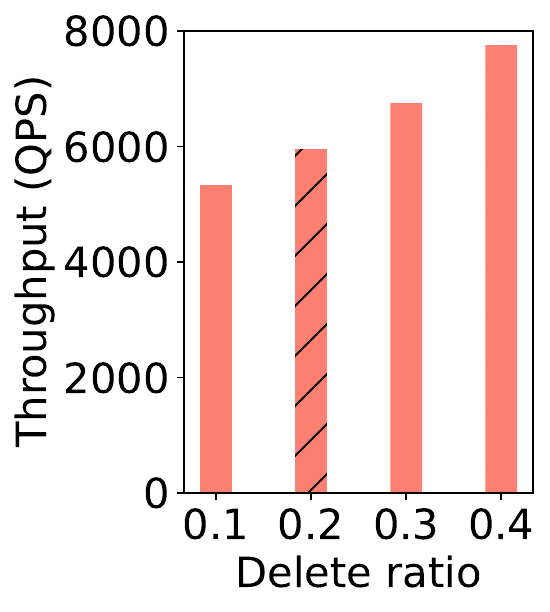}}
    \subfloat[{60\%R} \label{fig:delete-write0.4}]{\includegraphics[width=0.5\textwidth]{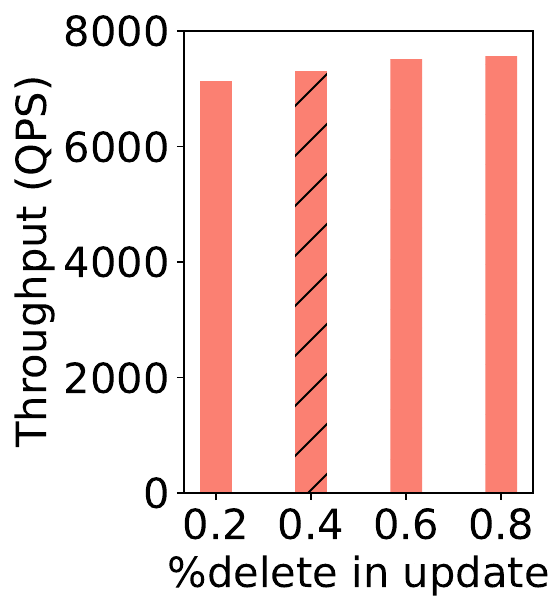}}
    \vspace{-0.3cm}
    \caption{{Performance under delete. (R: read, D: delete)}}
    \Description[short description]{long description}
    \label{fig:delete}
  \end{minipage}
  \vspace{-0.3cm}
\end{figure*}

\vspace{0.1cm}
\noindent{\textbf{Drift tolerance.}}
We prepare 1-million datasets derived from the ImageNet dataset. We reserve 1/10 categories for generating drift. We use a mixing ratio of vectors from the reserved categories and those from the original categories (not in the 1 million for index building) to create workloads with different drifts. The workloads consist of 4 batches of 200k vectors for insertion and 1k query vectors, such that both insertion and query exhibit drift. 
Figure~\ref{fig:ablation-drift} shows the recall and throughput as we insert data batches and then run ANN queries with a mixing ratio from 0 to 0.8. The $nprobe$ and $k'/k$ are selected to be the best search configuration with recall $\ge$ 0.99. The throughput descrease as more vectors are added resulting in more vectors to scan in each partition.
For search quality, we observe at this high recall, the recall improvement of training persists across different drifts. As more data are added the recall degrades slightly. The result showed the robustness of IVF and \nameindex{} training process against moderate drifts for embeddings from the same model.
We also evaluate on RSNET-2048 and observe similar results. 
For embeddings from different models or entirely distinct sources, we recommend building different indexes.

\vspace{0.1cm}
\noindent{\textbf{Decoupling index parameters for read and write.}}
We start with an index on 1 million vectors and select the $nprobe$ and $k'/k$ for recall $\ge 0.99$. We insert batches of 200k vectors and measure the recall at the same configuration. 
We derive the true nearest neighbor after each batch insert in prior. 
Figure~\ref{fig:ablation-decouple} shows the importance of separating the learned parameters for search from the parameters for insert. 
If the learned parameters are used to compress new vectors during insert, the recall drops. The reason is that only keeping learned parameters is inconsistent with our training scheme and the approximated similarity will not follow the expected distribution, as discussed in Section~\ref{subsec:index:discuss}. We observe in experiments that new vectors that are not nearest neighbors can have a higher approximated similarity than true neighbors, and the true neighbors in the added data can have significantly lower approximated similarity than those neighbors in the original dataset.

\vspace{0.1cm}
\noindent{\textbf{Deletion.}} We evaluate the index performance under deletion using the DPR-768 dataset and 32 clients. We select search parameters that achieve recall=0.99.
Figure~\ref{fig:all-delete} shows that for workloads involving both search and deletion, the throughput increases with the ratio of deletions. The trend is similar to the results in Figure~\ref{fig:index-con-rw} when the ratio of insert increases. The higher throughput is because insertion and deletion operations are cheaper than ANN search operations.
Figure~\ref{fig:delete-write0.4} shows that under workloads of 60\% search and 40\% of insert and delete, the throughput is only slightly higher when varying the ratio of delete, since insert operations calculate IVF assignment and compress the vectors.
Since we do not modify the coarse-grained partitioning when deleting the data, the recall can be maintained, as the close neighbors are likely to be selected from nearby partitions.

The Appendix contains additional results on full-range throughput-recall tradeoff, effects of Euclidean distance metric, deletion, ablation study for training and early termination.

\subsection{System Comparison}
\label{subsec:eval:sys}

We compare the performance of \namesys{} against the five distributed vector database baselines at 0.98 recall for $k=10$, using the two 10-million scale datasets.
For Cassandra, we use the same configuration for graph and beam search width during index construction.
However, since it uses quantized vectors instead of the original vectors, we adjust  $k$ to be larger than 10, such that if the refine stage is performed, it can reach the recall of $0.98$. 
Specifically, the system uses a quantized graph index to return a larger number of candidate vectors, which are then processed by a refine stage to achieve recall10@10 of 0.98. 
For \namesys{}, Sharded-HNSW, Weaviate, and Cassandra, we run one shard per node. For Milvus, we run a number of virtual
QueryNode according to the number of node settings used for other systems. The QueryNodes are evenly distributed among the physical nodes in a Kubernetes cluster. We use multiple distributed clients to saturate the systems, then report the peak throughputs.

\begin{figure}[t]
  \centering
  \includegraphics[width=0.48\textwidth]{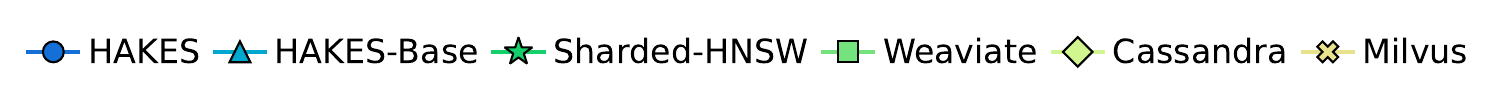}\\
  \vspace{-0.3cm}
  \subfloat[DPR-768-10m\label{fig:sphere-768-10m-vdb-node}]{\includegraphics[width=0.23\textwidth]{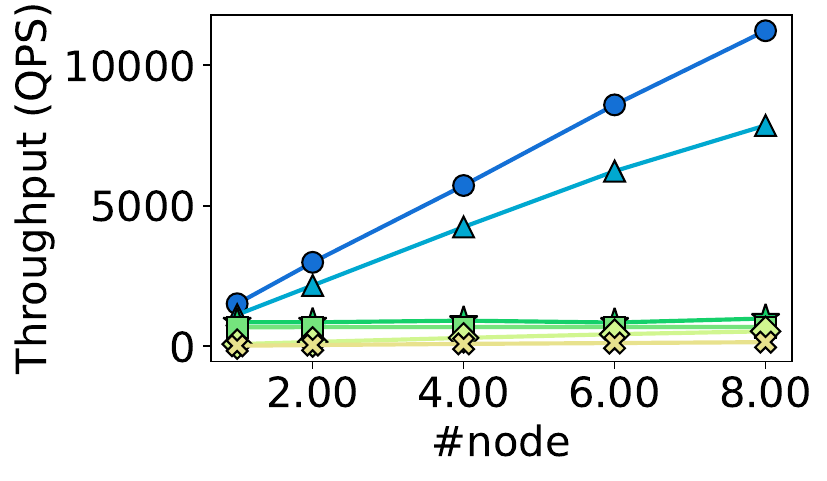}}
  \subfloat[E5-1024-10m\label{fig:sphere-1024-10m-vdb-node}]{\includegraphics[width=0.23\textwidth]{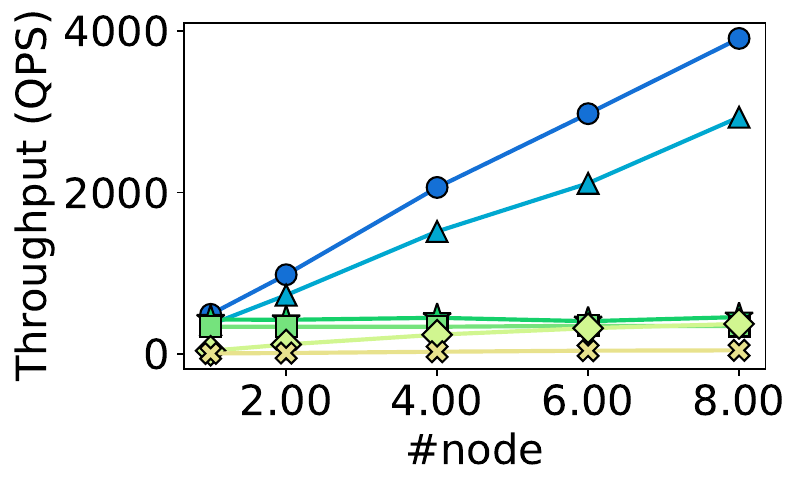}}
\vspace{-2mm}
  \caption{Scalability under read-only workload.}
  \Description[short description]{long description}
  \label{fig:dist-serv-node}
\end{figure}

\vspace{0.1cm}
\noindent\textbf{Scaling with the number of nodes.}
Figure~\ref{fig:dist-serv-node} compares the systems' throughputs with varying numbers of nodes. It can be seen that
\namesys{} and \namesys{}-Base scale linearly because the load of both the filtering and refinement stages are distributed evenly across
the nodes. The filtering stage of concurrent requests can be processed at different nodes in parallel. In Weaviate, a
request is sent to all the shards. Although the graph index size and the number of vectors in each shard decrease with more
nodes, the search cost at each shard does not decrease linearly. This is consistent with the results of ELPIS in Section~\ref{subsec:eval:index}, confirming that graph indexes do not scale well by partitioning. 
Sharded-HNSW achieves slightly better throughput than Weaviate, but the same trend is observed.
Milvus' throughput increases with the number of read shards, due to the reduced read load. However, the small read shard size of 1GiB leads to  a large number of subgraphs (over 20 for Sphere-768-10m and over 30 for
Sphere-1024-10m), all of which need to be searched, meaning that the throughputs are low. 
In Cassandra, a single node contains multiple shards, the number of which is affected by its Log-structured merge (LSM) tree compaction process.
We observe that the number of shards per node decreases as the number of nodes increases, which explains the increasing throughput. 
At 8 nodes, there is one shard per node and the performance is similar to that of Weaviate and Sharded-HNSW.
The improvement of  \namesys{} over \namesys{}-Base shows the benefit of \nameindex{} in reducing the search cost with its learned compression and optimizations. 

\begin{figure}[t]
  \centering
   \includegraphics[width=0.48\textwidth]{arxiv-figures/legends/legends-for-6.pdf}\\
   \vspace{-0.5cm}
  \subfloat[DPR-768-10m\label{fig:sphere-768-10m-vdb-rw}]{\includegraphics[width=0.23\textwidth]{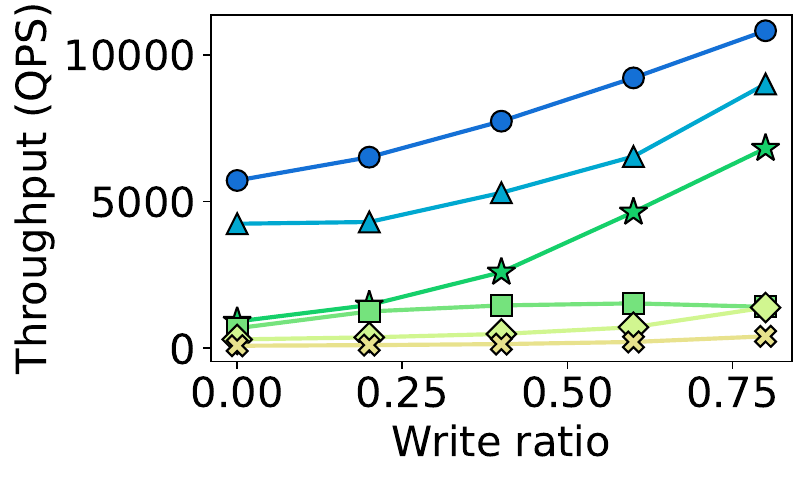}}
  \subfloat[E5-1024-10m\label{fig:sphere-1024-10m-vdb-rw}]{\includegraphics[width=0.23\textwidth]{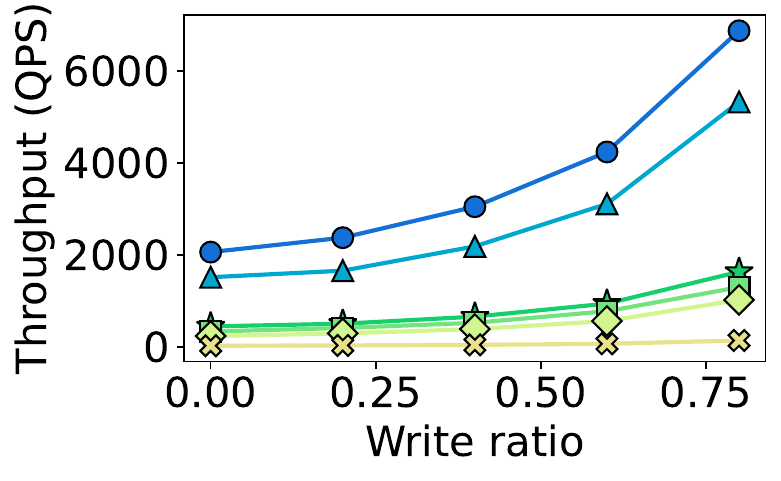}}
  \vspace{-2mm}
  \caption{Throughput under read-write workload (4 nodes).}
  \Description[short description]{long description}
  \label{fig:dist-serv-rw}
\end{figure}

\vspace{0.1cm}
\noindent\textbf{Performance under concurrent read-write workloads.} We fix 4 nodes for all systems
and vary the write ratio. Figure~\ref{fig:dist-serv-rw} shows that all systems have higher throughput as the write ratio
increases. For Weaviate, Sharded-HNSW, and Cassandra, the write request is only processed by one shard, as opposed to by
all the shards for read requests. Sharded-HNSW has the highest performance among baselines that use graph-based indexes, due to its
C++ implementation. \namesys{} and \namesys{}-Base outperform all the other baselines by a considerable margin, and \namesys{} has higher throughputs than \namesys{}-Base. Even though each write request
needs to be processed by all IndexWorkers,  \namesys{} is more efficient than the others in processing the write request, because it only computes the quantized vector and updates the IVF structure. In the other baselines, each node has to perform a read to identify neighbor vectors and network edges to be updated.

\section{Related Work} \label{sec:related}

\noindent{\bf Managing ANN index update.} 
Graph indexes, like HNSW~\cite{hnsw} and LVQ~\cite{lvq}, rebuild the graph connections locally. SPFresh~\cite{spfresh} uses an in-memory graph to index a large number of partitions of vectors on disk and proposes a scheme to keep the partition size small for stable serving latency.
At the system level, sharding is employed to reduce the impact of inserting new vectors~\cite{lanns, manu}. \nameindex{} appends vectors to the partitioning-based index and uses tombstone for deletion to minimize interference on search and maintain the high recall without changing the search configurations. The low read-write contention allows \namesys{} to maintain replicated global indexes for better scaling performance. 

\vspace{0.1cm}
\noindent{\bf Adaptive query search.}
Several works exploit characteristics of the queries and the immediate search results to improve the vector search.
Auncel~\cite{auncel}, iDistance~\cite{idistance}, and VBASE~\cite{vbase} leverage precise similarity scores to determine if a search can terminate early, making them unsuitable for the \nameindex{}'s filter stage that operates on compressed vectors.
ADSampling~\cite{adsampling} progressively uses more dimensions to compare vector pairs.
Learning-based methods like LEQAT~\cite{leqat, laet} employ predictive models that incur costly training and inference overhead. 
In contrast, \nameindex{}'s early termination check is lightweight as it is based on simple computation over statistics available during search. 

\noindent{\bf Vector quantization.} 
ScaNN~\cite{scann}, SOAR~\cite{soar}, and QUIP~\cite{quip} learn quantization codebooks to reduce the error of the approximation of the inner product. 
RabitQ~\cite{rabitq} quantizes vectors into binary representations and provides a theoretical error bound on the similarity score. 
OPQ~\cite{opq} and DOPQ~\cite{dopq} learn data transformation and quantization codebooks to reduce the error in reconstructing the original vectors. 
~\cite{spreading} learns a transformation matrix to spread out vectors for quantized code assignment and keep the top k neighbors close. These works have a different optimization objective from ours. In particular, we learn the dimensionality reduction and product quantization together to reduce the local similarity score distribution mismatch.  
Other works from the information retrieval community~\cite{distill-vq, repconc, mevi} propose to jointly train embedding models and product quantization codebooks, which is similar to our approach. However, their training objective is to capture the semantic similarity between data vectors, which requires access to the embedding model or labels on semantic relevance for the original data. Our approach does not require such access. 

\vspace{-2mm}
\section{conclusion}
\label{sec:conclusion}

We presented a scalable, distributed vector database \namesys{} that
supports approximate nearest neighbor search with high recall and
high throughput for online services that are subject to concurrent
read-write workloads. HAKES employs a novel partitioning-based
index that adopts a two-stage process with learned compression
parameters. We proposed a lightweight training process and a separation
of index parameters to support vector insert. HAKES adopts
a disaggregated architecture specifically designed to exploit the
access pattern of the new index. We compared HAKES against existing
distributed vector databases, showing that our system achieves
up to 16× throughputs over the baselines at high recall regions.


\bibliographystyle{ACM-Reference-Format}
\bibliography{ref}

\clearpage
\appendix
\section{Appendix}

\subsection{Index Building and Search Configurations}

We build each index with different configurations. Then we select the best configuration for throughput-recall trade-off for high recall $\ge 0.9$ to report in Figure~\ref{fig:index-recall}. The explored configurations follow the suggestions in the respective original publications, documentation in their code repositories, or established index benchmarks~\cite{ann-benchmarks}. Specifically, IVF-based indexes use $nlist \in \{1024, 2048, 4096\}$ and $nprobe \in \{1, 5, 10, 50, 100, 200, 300, 400, 500\}$. For PQ, $nbits=4$ to leverage the fast scan implementation, and $m \in \{d/2, d/4, d/8\}$. For OPQ, the output dimension $d_r \in \{d, d/2, d/4, d/8\}$. ScaNN and SOAR are built with $dimensions\_per\_block = 2, T = 0.2$, $\#leaves \in \{1024, 2048, 4096\}$ and SOAR additionally set $overretrieve\_factor = 2$ and $\lambda = 1.5$. RabitQ uses recommended settings, $C = 4096, B_q=4$, and searches with the same $nporbe$ value range above. HNSW is built with $M \in \{16, 32, 64\}$ and $ef\_construction \in 200$, ELPIS with $max\_leaf\_size = N/ 5, M = 32, ef\_construction = 200$, and LSH-APG with $L=2, K=18, M = 24, ef\_construction = 200$. The three graph indexes use search parameters $ef\_search \in \{128, 256, ..., 4096\}$. LVQ is built with the three modes $Vamana, LVQ8, LVQ4\times8$, window search size $200$ and $\alpha=0.95, R=\{32, 64, 128\}$.
Falconn++ use build configurations $k = 2, L \in \{100, 200, 400, 800\}, iprobe \in \{2,4,6,8\}$ and search configurations $qprobe \in \{2000, 4000, 38000\}$. For LCCS, $L\in \{64, 128, ..., 2048\}$ and the search setting values are in $\{1,2,4, ..., 512\}$.
When applicable, the refine stage with exact distance is done with $k'/k = \{1, 10, 20, 50, 100, 200, 300, 400, 500\}$.
For \nameindex{} search optimizations, we always enable IVF centroid quantization and use $t=k'/200, n_t=10$, even though we note tunning the early termination parameters for each dataset can yield even better throughput-recall trade-off.

The selected build configuration variable values are presented in Table~\ref{tab:app:build-config}. Indexes using one recommended build configuration and other configuration parameters using one recommended value are not included in the table.

\begin{table*}[t]
    \centering
    \caption{Selected index building configuration values for index benchmarking.}
    \label{tab:app:build-config}
    \vspace{-0.3cm}
    \begin{tabular}{|c|c|c|c|c|c|c|}
        \hline
         & Variables & DPR-768 & OPENAI-1536 & MBNET-1024 & RSNET-2048 & GIST-960 \\ \hline
         IVF & $nlist$ & $1024$ & $1024$ & $2048$ & $2048$ & $1024$ \\ \hline
         IVFPQ\_RF & $nlist, m$ & $1024,d/2$ & $1024,m=d/8$ & $1024,d/4$ & $1024,d/8$ &  $1024,d/2$  \\ \hline
         OPQIVFPQ\_RF & $dr,m,nlist$ & $d/4,d/2,1024$ & $d/8,d/2,1024$ & $d/4,d/2,2048$ & $d/8,d/2,1024$ & $d/4,d/2,1024$  \\ \hline
         HNSW & M & $32$ & $32$ & $32$ & $16$ & $32$ \\ \hline
         ScaNN & $\#leaves$ & $4096$  &$4096$  & $2048$ & $4096$ & $4096$ \\ \hline
         SOAR & $\#leaves$ & $4096$  & $2048$ &$4096$  & $4096$  & $4096$\\ \hline
         Falconn++ & $iprobe, L$ & $16, 400$ & $16, 100$ & $16, 200$ & $16, 100$ & $8, 400$ \\ \hline
         LCCS & $L$ & $2048$  & $2048$ & $2048$ & $2048$ & $2048$ \\ \hline
         LVQ & $mode, R$ & $LVQ4\time 8,64$ & $LVQ4\time 8,64$ & $LVQ4\time 8,64$ & $LVQ4\time 8,64$ & $LVQ4\time 8,64$ \\ \hline
         \nameindex{} & $dr,m,nlist$ & $d/4,d/2,1024$ & $d/8,d/2,1024$ & $d/4,d/2,2048$ & $d/8,d/2,1024$ & $d/4,d/2,1024$  \\ \hline
    \end{tabular}
\end{table*}

\subsection{Full Range Throughput-recall Trade-off}

\begin{figure*}[t]
  \centering
  \includegraphics[width=0.98\textwidth]{arxiv-figures/legends/legends.pdf}\\
  \vspace{-0.3cm}
  \subfloat[DPR-768\label{fig:sphere-768-1m-recall}]{\includegraphics[width=0.195\textwidth]{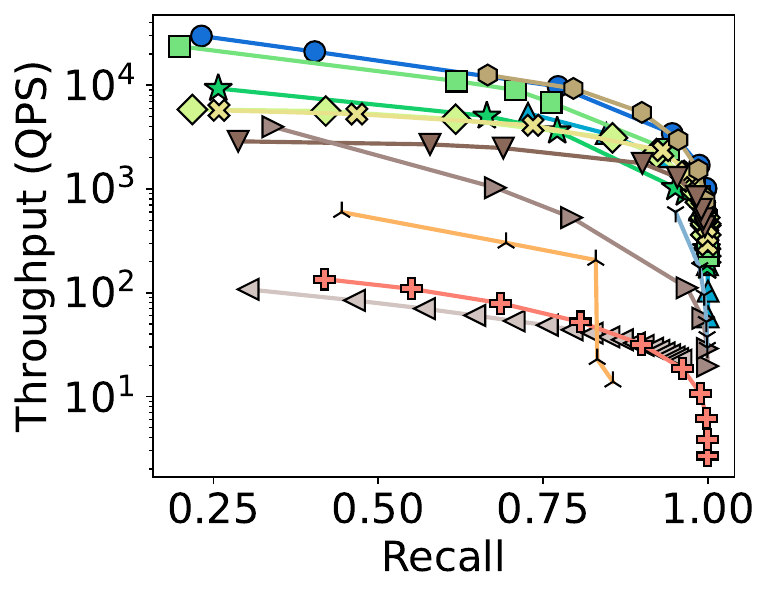}}
  \subfloat[OPENAI-1536\label{fig:DBpedia-openai-recall}]{\includegraphics[width=0.195\textwidth]{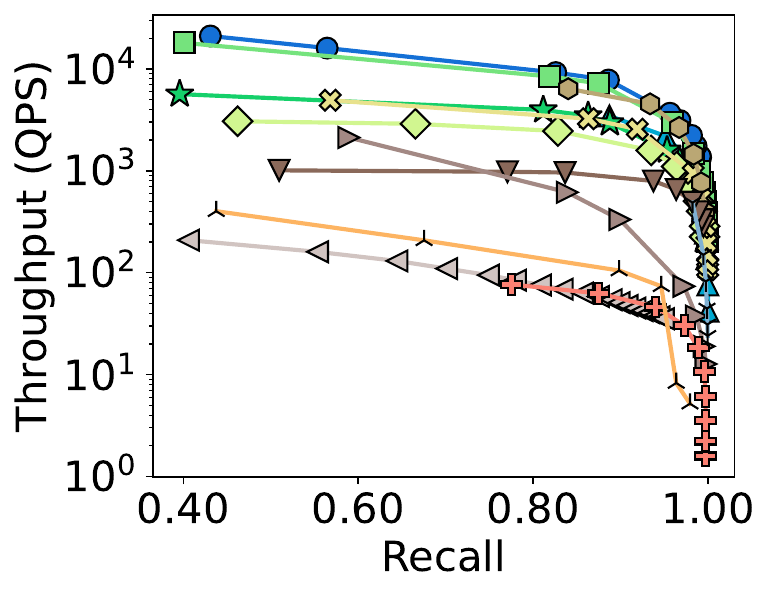}}
  \subfloat[MBNET-1024\label{fig:mbnet-1024-recall}]{\includegraphics[width=0.195\textwidth]{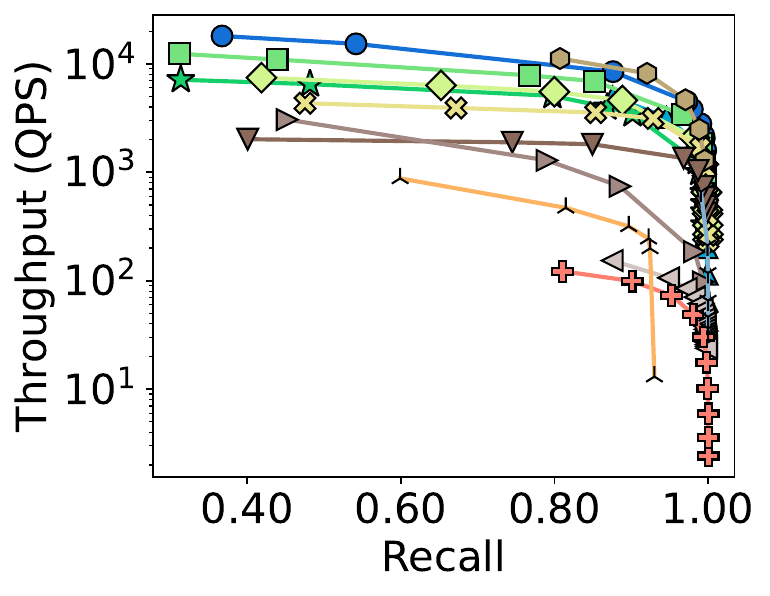}}
  \subfloat[RSNET-2048\label{fig:rsnet-2048-recall}]{\includegraphics[width=0.195\textwidth]{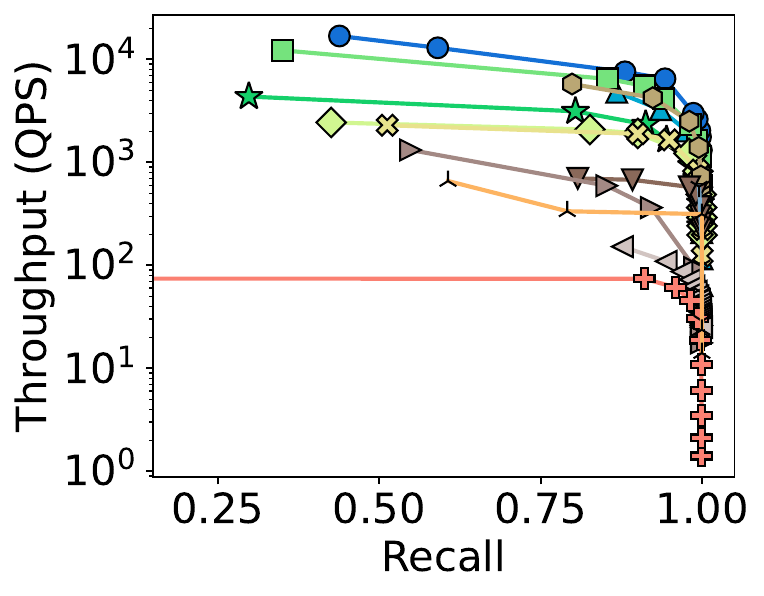}}
  \subfloat[GIST-960\label{fig:gist-960-recall}]{\includegraphics[width=0.195\textwidth]{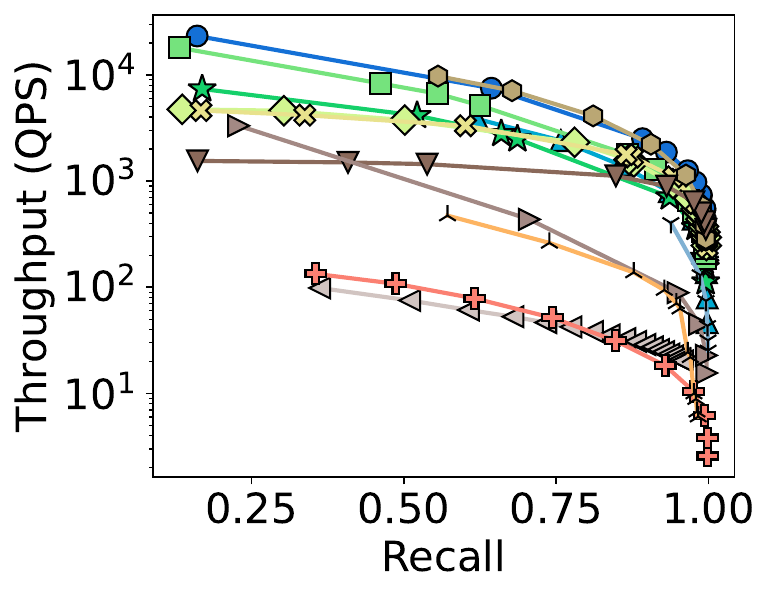}}
  \vspace{-0.3cm}
  \caption{Throughput vs. recall for sequential reads.}
  \label{fig:index-recall-full}
  \Description[short description]{long description}
\end{figure*}

Figure~\ref{fig:index-recall-full} shows the throughput-recall trade-off on the full recall range based on the best configuration settings tested. \nameindex{} achieves competitive performance over the whole recall range, even though \nameindex{} targets the high-recall range and the learned search index parameters are designed to focus on minimizing the similarity score distribution in the vicinity of query vectors.

\subsection{Recall Improvement on 1M Datasets}

Table~\ref{tab:app:index-enhance} presents the result on the 1 million datasets. The self-supervised learning process of \nameindex{} enhances the recall in the majority of the configurations in the high recall region. When $nprobe$ is low, the IVF partition ranking due to learned dimension reduction may leave out those that contain true nearest neighbors, the effect becomes negligible as $nprobe$ is large.

\begin{table*}[t]
    \centering
    \caption{Recall improvement at different search configurations (1-million scale datasets).}
    \label{tab:app:index-enhance}
    \vspace{-0.3cm}
\resizebox{\textwidth}{!}{
    \begin{tabular}{|c|c|ccc|ccc|ccc|ccc|}
    \hline
    \multicolumn{2}{|c|}{IVF $nprobe$ (total 1024)}  & \multicolumn{3}{c|}{10} & \multicolumn{3}{c|}{50} & \multicolumn{3}{c|}{100} & \multicolumn{3}{c|}{200}\\
    \hline
    \multicolumn{2}{|c|}{$k'/k$} & 10 & 50 & 200 & 10 & 50 & 200 & 10 & 50 & 200 & 10 & 50 & 200 \\ \hline
    \multirow{2}{*}{DPR-768} 
        & Base & 0.708 & 0.790 & \textbf{0.799} & 0.795 & 0.940 & 0.966 & 0.801 & 0.957& 0.988 & 0.803 & 0.962 & 0.995 \\ 
        & Learned    & \textbf{0.772} & \textbf{0.798} & 0.798 & \textbf{0.907} & \textbf{0.964} & \textbf{0.968} & \textbf{0.922} & \textbf{0.987} & \textbf{0.991} & \textbf{0.927} & \textbf{0.994} & \textbf{0.999}\\ \hline
    \multirow{2}{*}{OPENAI-1536} 
        & Base & 0.875 & 0.895 & \textbf{0.897} & 0.940 & 0.969 & 0.972 & 0.952 & 0.983 & 0.987 & 0.957 & 0.990 & 0.994 \\ 
        & Learned    & \textbf{0.886} & \textbf{0.897} & \textbf{0.897} & \textbf{0.956} & \textbf{0.972} & \textbf{0.974} & \textbf{0.969} & \textbf{0.986} & \textbf{0.987} & \textbf{0.975} & \textbf{0.994} & \textbf{0.995}\\
    \hline
    \multirow{2}{*}{MBNET-1024} 
        & Base & 0.852 & \textbf{0.886} & \textbf{0.886} & 0.930 & \textbf{0.980} & \textbf{0.981} & 0.938 & 0.991& \textbf{0.993} & 0.940 & 0.995 & \textbf{0.997} \\ 
        & Learned & \textbf{0.879} & 0.884 & 0.884 & \textbf{0.972} & \textbf{0.980} & 0.980 & \textbf{0.983} & \textbf{0.992} & 0.992 & \textbf{0.986} & \textbf{0.996} & 0.996\\
    \hline
    \multirow{2}{*}{RSNET-2048} 
        & Base & 0.911 & \textbf{0.948} & \textbf{0.949} & 0.947 & 0.992 & \textbf{0.993} & 0.951 & 0.997& \textbf{0.998} & 0.952 & 0.998 & \textbf{0.999} \\ 
        & Learned & \textbf{0.943} & \textbf{0.948} & 0.948 & \textbf{0.986} & \textbf{0.993} & \textbf{0.993} & \textbf{0.980} & \textbf{0.998} & \textbf{0.998} & \textbf{0.992} & \textbf{0.999} & \textbf{0.999}\\ 
    \hline
    \multirow{2}{*}{GIST-960} 
        & Base & 0.555 & 0.679 & \textbf{0.702} & 0.645 & 0.867 & 0.936 & 0.651 & 0.884& 0.967 & 0.652 & 0.888 & 0.975 \\ 
        & Learned & \textbf{0.647} & \textbf{0.689} & 0.690 & \textbf{0.828} & \textbf{0.932} & \textbf{0.945} & \textbf{0.847} & \textbf{0.966} & \textbf{0.984} & \textbf{0.852} & \textbf{0.975} & \textbf{0.995}\\
    \hline
    \end{tabular}
    }
\end{table*}

\subsection{Early Termination Parameters}

\begin{figure}[t]
  \centering
  \includegraphics[width=0.47\textwidth]{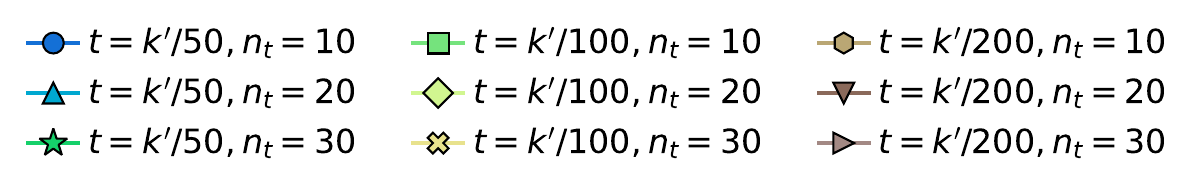}\\
  \vspace{-0.5cm}
  \subfloat[DPR-768\label{fig:dpr-768-ab-et}]{\includegraphics[width=0.242\textwidth]{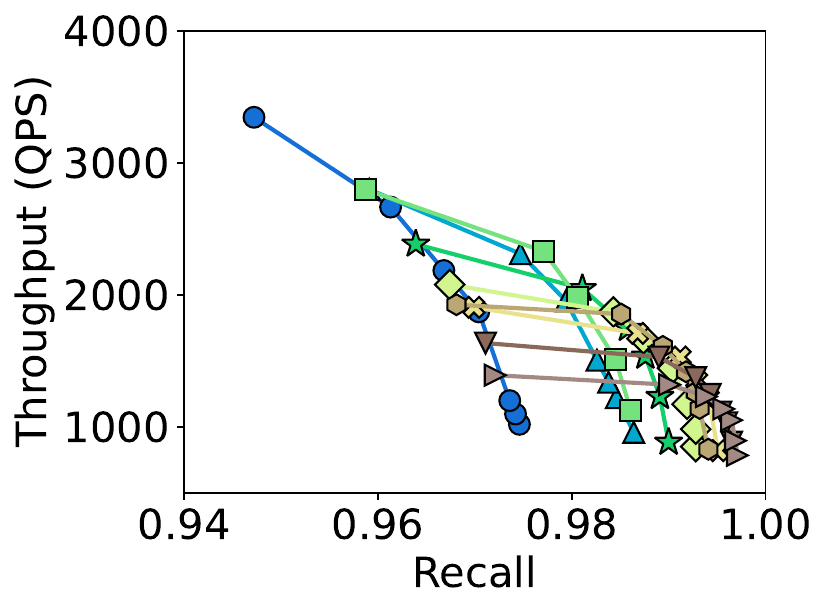}}
  \subfloat[GIST-960\label{fig:gist-960-ab-et}]{\includegraphics[width=0.235\textwidth]{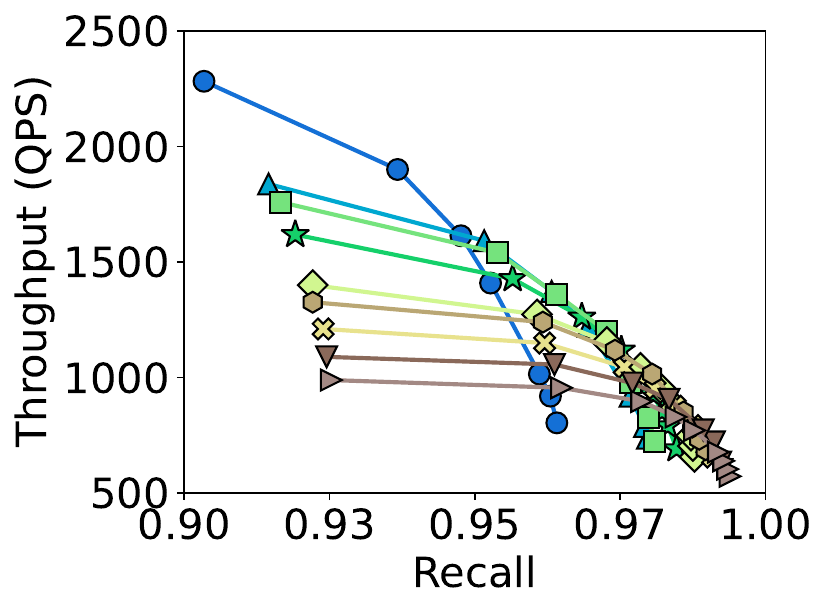}}
  \caption{Effect of early termination parameters.}
  \Description[short description]{long description}
  \label{fig:ablation-et}
\end{figure}

Setting the same $nprobe$ to achieve high recall causes all queries to scan a large number of partitions according to the harder ones, even though some of the queries can identify all nearest neighbors in the first few partitions. As discussed in Section~\ref{subsec:index:optim}, \nameindex{} introduces an early termination checking to adapt the search process of a query based on its intermediate results, with two parameters, $n_t$ and $t$. In our implementation, $t$ is set to a value relative to $k'$, because for a larger $k'$ the intermediate result set keeps further away vectors and a partition adds more candidates. Figure~\ref{fig:ablation-et} shows the effect of these two parameters on DPR-768 and GIST-960. We fix $nprobe=200$ so that $recall > 99.5\%$ on the two datasets if query vectors are evaluated against all vectors in their closest 200 partitions with exact similarity calculation. Each curve is plotted by varying $k'/k$. 
The smaller $t$ and $n_t$, the earlier the criterion is satisfied, resulting in higher throughput.
However, low  $t$ and $n_t$ can lead to queries terminating prematurely and low recall.  

\begin{figure}[t]
  \centering
  \includegraphics[width=0.3\textwidth]{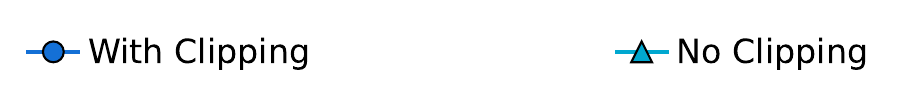}\\
  \vspace{-0.5cm}
  \subfloat[DPR-768\label{fig:dpr-768-ab-et-clip}]{\includegraphics[width=0.242\textwidth]{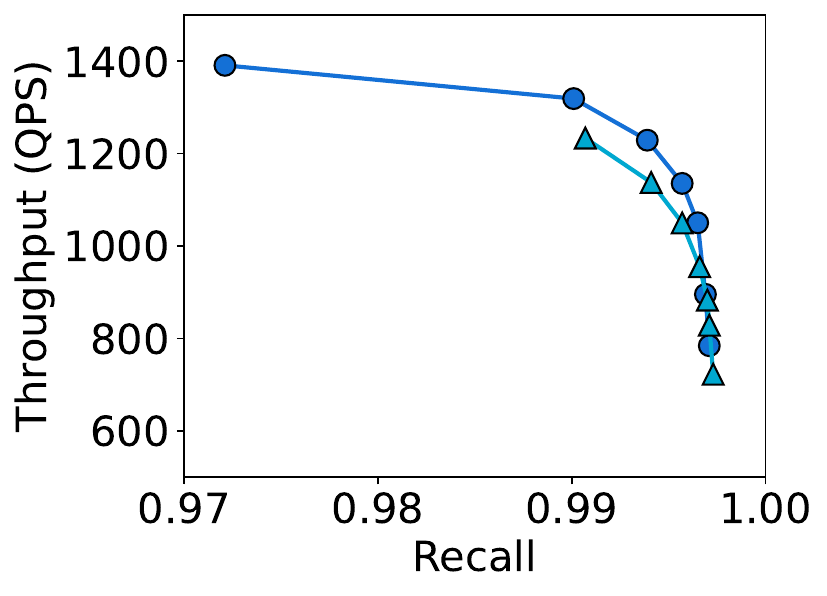}}
  \subfloat[GIST-960\label{fig:gist-960-ab-et-clip}]{\includegraphics[width=0.235\textwidth]{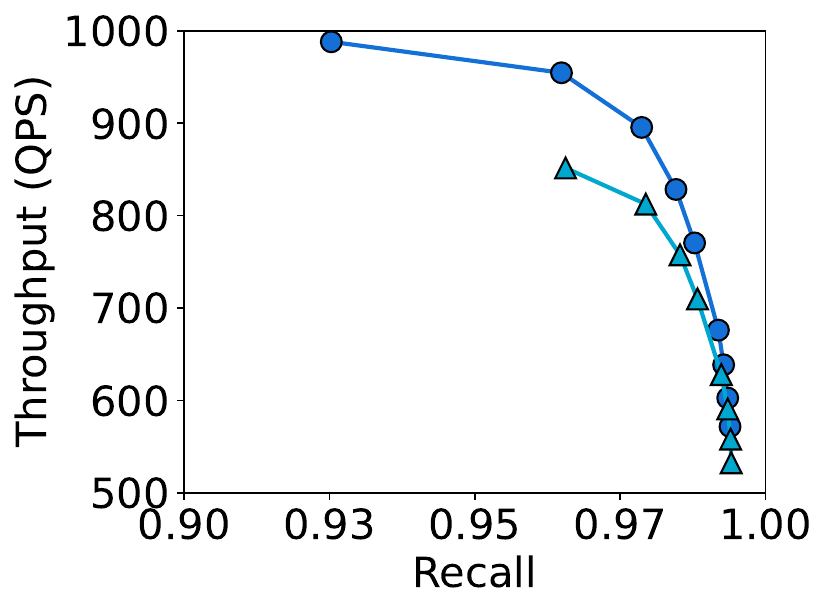}}
  \caption{Effect of stopping at \textit{nprobe} with early termination.}
  \Description[short description]{long description}
  \label{fig:ablation-et-clip}
\end{figure}

We then study the effect of leaving out $nprobe$ and using only the termination criteria defined by $t$ and $n_t$. We set $t = k'/200$ and $n_t = 30$. Figure~\ref{fig:ablation-et-clip}, shows that if we do not clip the search with $nprobe$ the throughput-recall trade-off is worse. In \nameindex{} filter stage, the similarity approximation is lossy and so that partitions further away can still add candidates to the $k'$ vector intermediate results. For some queries, $n_t$ can only be satisfied after many more partitions than needed are scanned, leading to a drop in throughput. Therefore, \nameindex{} uses both criteria and terminates the query when it satisfies one.    

\subsection{Drift Tolerance Result on RSNET-2048}

\begin{figure}[t]
  \centering
  \includegraphics[width=0.46\textwidth]{arxiv-figures/exp-drift/legends-drift.pdf}\\
  \vspace{-0.5cm}
   \subfloat[Recall\label{fig:drift-recall-resnet}]{\includegraphics[width=0.46\textwidth]{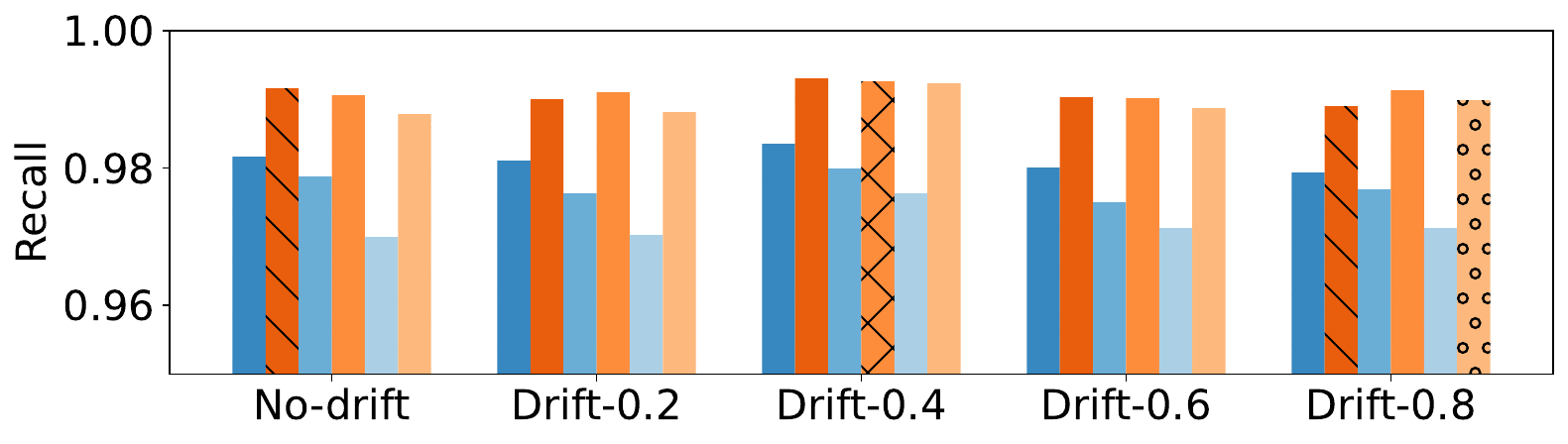}}\\
  \vspace{-0.35cm}
  \subfloat[Throughput\label{fig:drift-throughput-resnet}]{
  \includegraphics[width=0.46\textwidth]{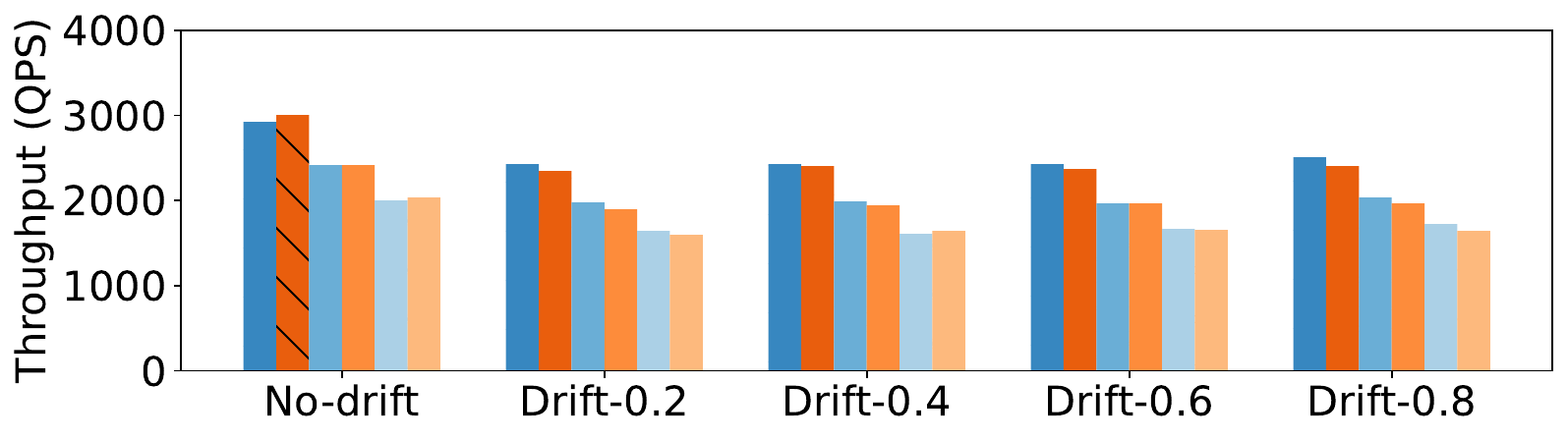}
  }
  \vspace{-0.3cm}
  \caption{Tolerance against data drift (RSNET-2048).}
  \label{fig:ablation-drift-resnet}
  \Description[short description]{long description}
\end{figure}

Similar to Figure~\ref{fig:ablation-drift}, Figure~\ref{fig:ablation-drift-resnet} shows the tolerance of drift on the other ImageNet embeddings where we have access to the true categories. We have similar observations that the recall improvement persists across different drifts at this high recall. The addition of the new vectors with drift can impact the recall for \nameindex{} but not more significantly than using the base parameters for search.

\subsection{Number of Near Neighbor (NN) in Learning}
\begin{figure}[t]
  \centering
  \includegraphics[width=0.47\textwidth]{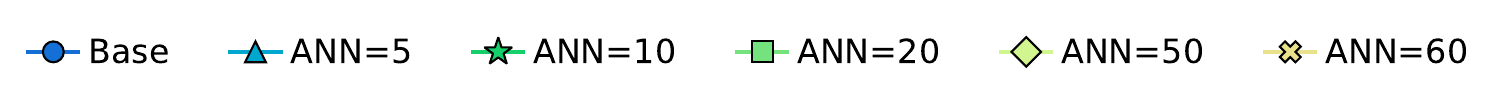}\\
  \vspace{-0.5cm}
  \subfloat[DPR-768-10m\label{fig:sphere-768-10m-nn}]{\includegraphics[width=0.242\textwidth]{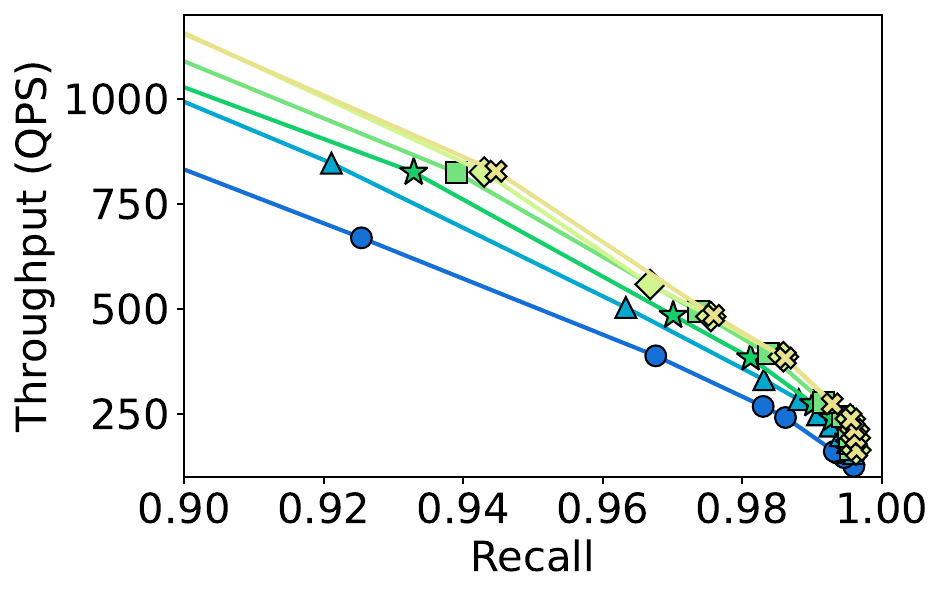}}
  \subfloat[E5-1024-10m\label{fig:sphere-1024-10m-nn}]{\includegraphics[width=0.235\textwidth]{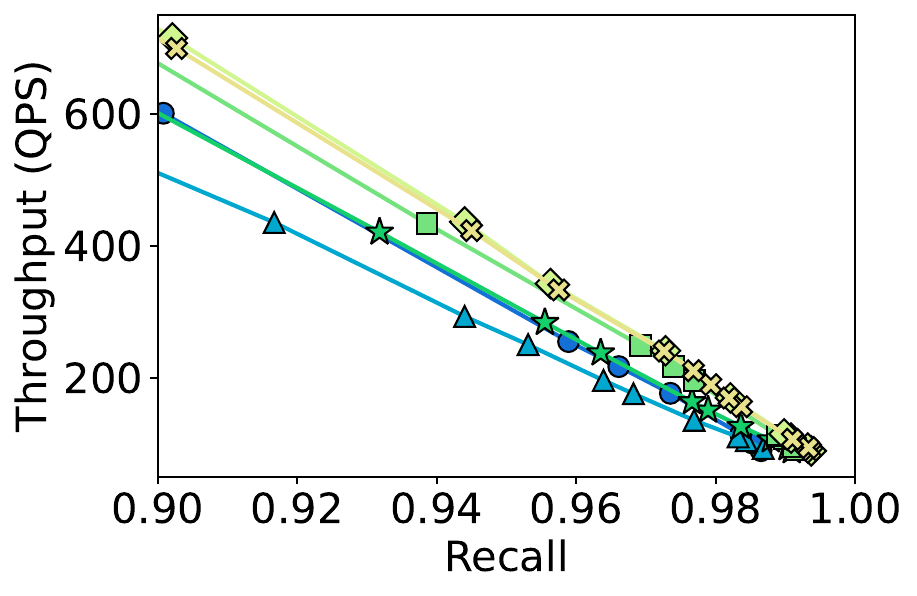}}
  \vspace{-0.3cm}
  \caption{Effect of \#ANN in training.}
  \Description[short description]{long description}
  \label{fig:ablation-nn}
\end{figure}

We plot the throughput-recall curve of \indexname{} with varying numbers of approximated nearest neighbors (NN) used in training.
Figure~\ref{fig:ablation-nn} shows that increasing NN improves the performance-recall curve, until around 50.
The result suggests that the training needs a sufficient number of vectors in close proximity.
The intuition is that the set of candidate vectors is larger than the set of final $k$ results. 
Therefore the ranking order of vectors in a larger region around the query is important to reduce the chance of missing true nearest neighbors when returning the candidate vectors.

\subsection{ Memory Usage in Distributed Serving}

\begin{figure}[t]
  \centering
  \includegraphics[width=0.47\textwidth]{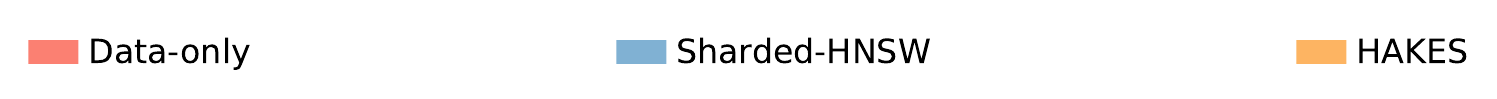}\\
  \vspace{-0.5cm}
  \subfloat[DPR-768-10m\label{fig:storage-spread-768}]{\includegraphics[width=0.235\textwidth]{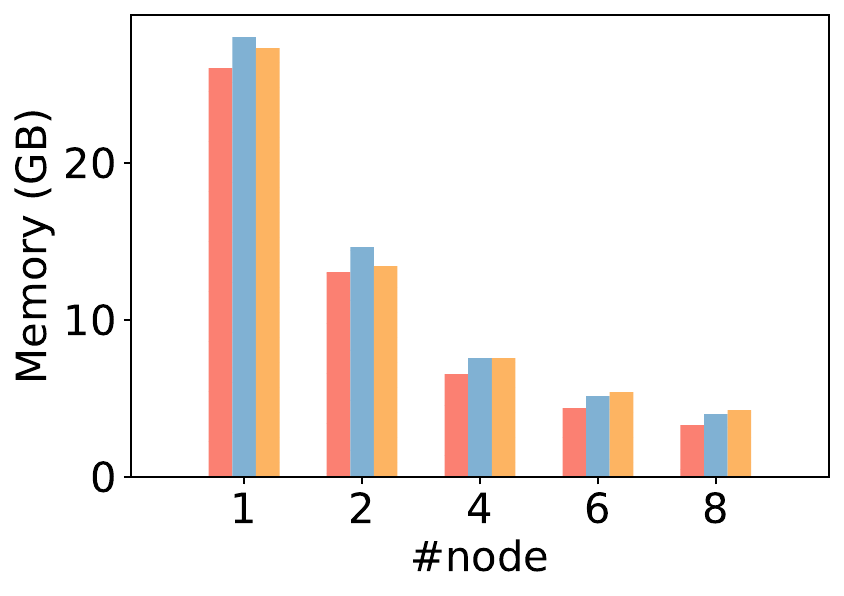}}
  \subfloat[E5-1024-10m\label{fig:storage-spread-1024}]{\includegraphics[width=0.235\textwidth]{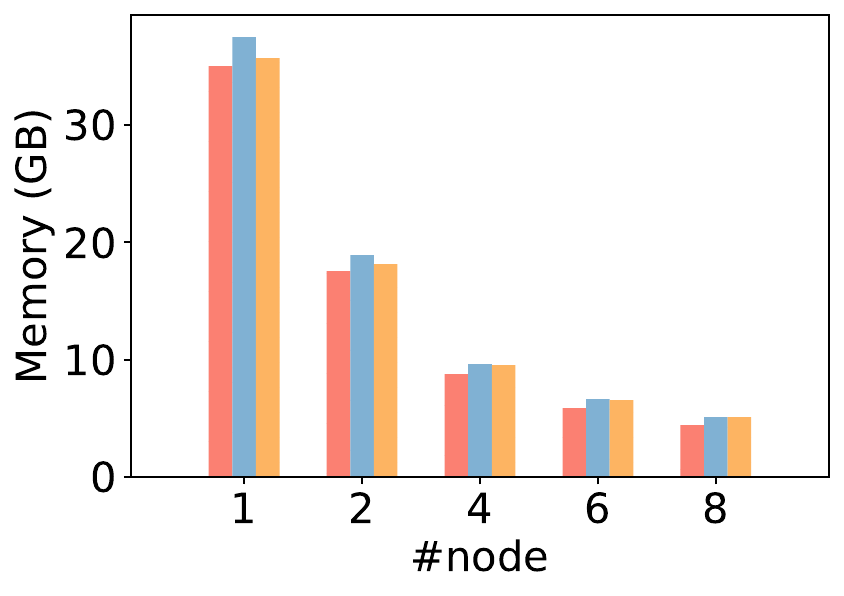}}
  \vspace{-3mm}
  \caption{Memory consumption per node.}
  \Description[short description]{long description}
  \label{fig:storage-spread}
\end{figure}

For high-dimension embedding vectors, the main storage cost comes from storing the vectors themselves. Replicating the compressed index in \namesys{} in the distributed setting has shown a significant throughput advantage compared to the small memory overhead it causes.
Figure~\ref{fig:storage-spread} shows the average memory consumption per node with varying number of nodes. It can be
seen that the memory overhead over the raw vectors is small for both \namesys{} and Sharded-HNSW across different node settings. In deployment, the number of IndexWorkers shall be approximated by dividing the overall throughput requirement by individual IndexWorker's achievable throughput and the number of RefineWorkers approximated based on the dataset size over the memory space of a single server. 

\subsection{Evaluation on Euclidean Distance}

\begin{figure*}
    \centering
    \begin{minipage}[c]{0.395\textwidth}
        \centering
        \includegraphics[width=\textwidth]{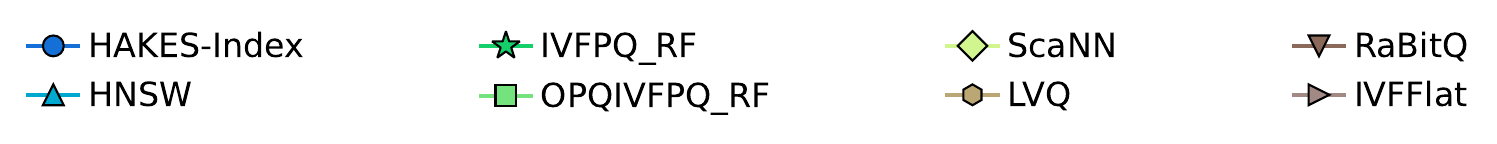}
        \vspace{-0.9cm}
        \\
        \subfloat[DPR-768\label{fig:dpr-768-l2}]{\includegraphics[width=0.5\textwidth]{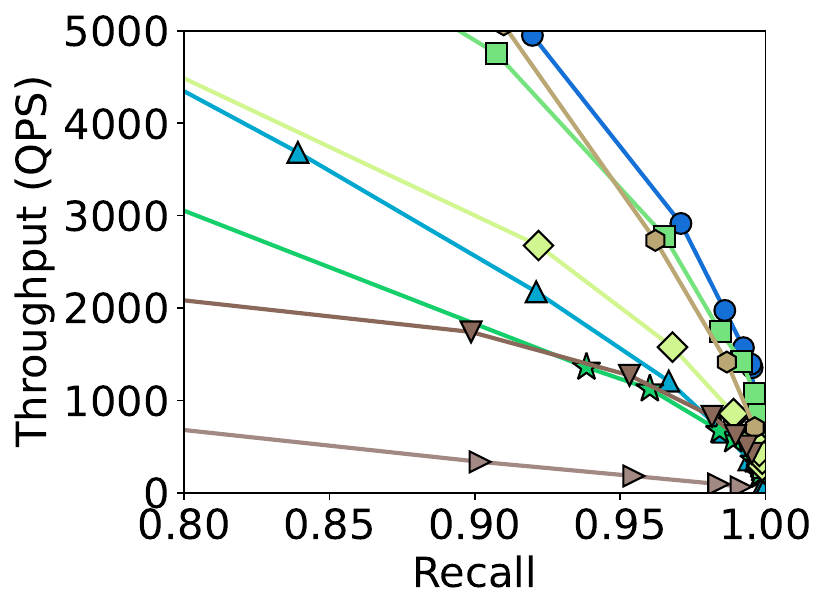}}
        \subfloat[GIST-960\label{fig:gist-960-l2}]  {\includegraphics[width=0.5\textwidth]{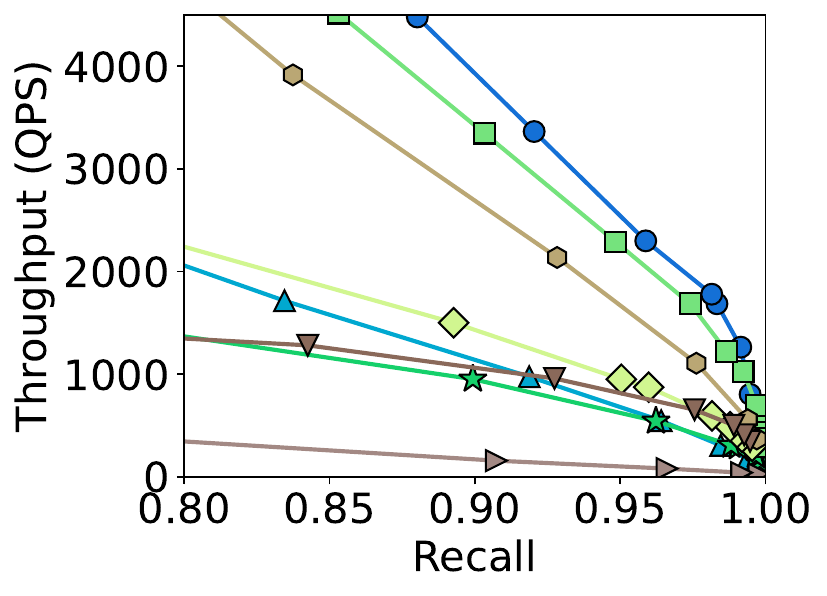}}
        \vspace{-0.3cm}
        \caption{Evaluation on Euclidean distance.}
        \label{fig:l2}
    \end{minipage}
    \begin{minipage}[c]{0.395\textwidth}
      \centering
      \includegraphics[width=\textwidth]{arxiv-figures/exp-l2/legends-l2.pdf}
      \vspace{-0.9cm}
      \\
      \subfloat[SIFT-128\label{fig:sift-128}]  {\includegraphics[width=0.5\textwidth]{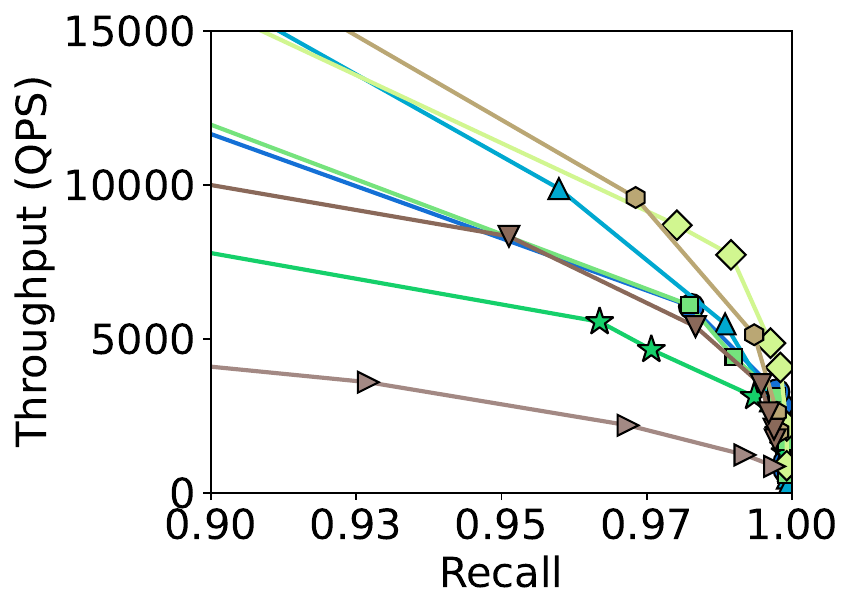}}
      \subfloat[DEEP-256\label{fig:deep-256}]{\includegraphics[width=0.5\textwidth]{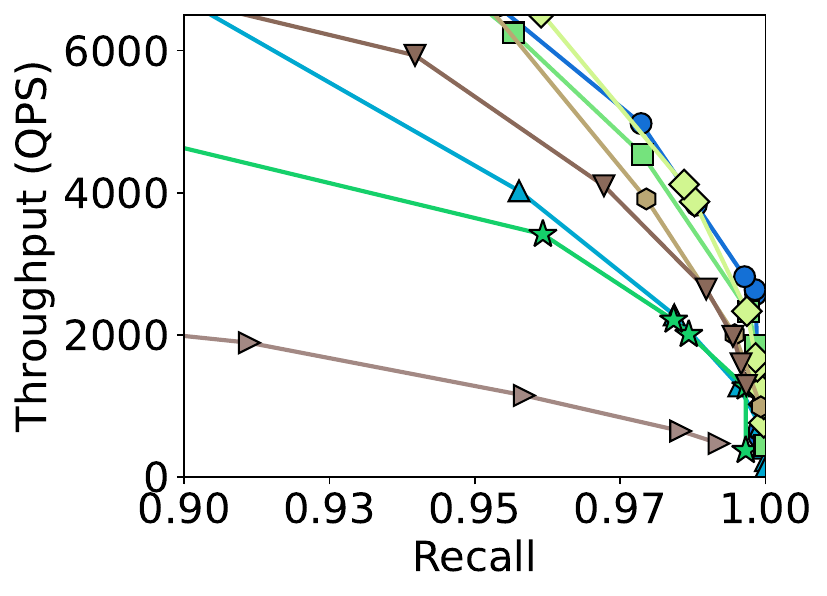}}
      \vspace{-0.3cm}
      \caption{Low-dimensional datasets.}
      \Description[short description]{long description}
      \label{fig:low-d}
    \end{minipage}
    \begin{minipage}[c]{0.20\textwidth}
        \centering
        \includegraphics[width=\textwidth]{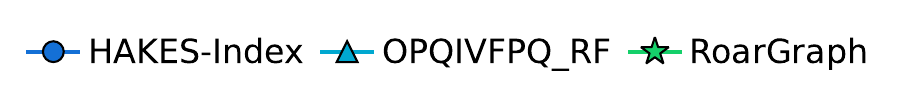}
        \vspace{-0.8cm}
        \\
        \subfloat[Text-to-image\label{fig:t2i}]{\includegraphics[width=\textwidth]{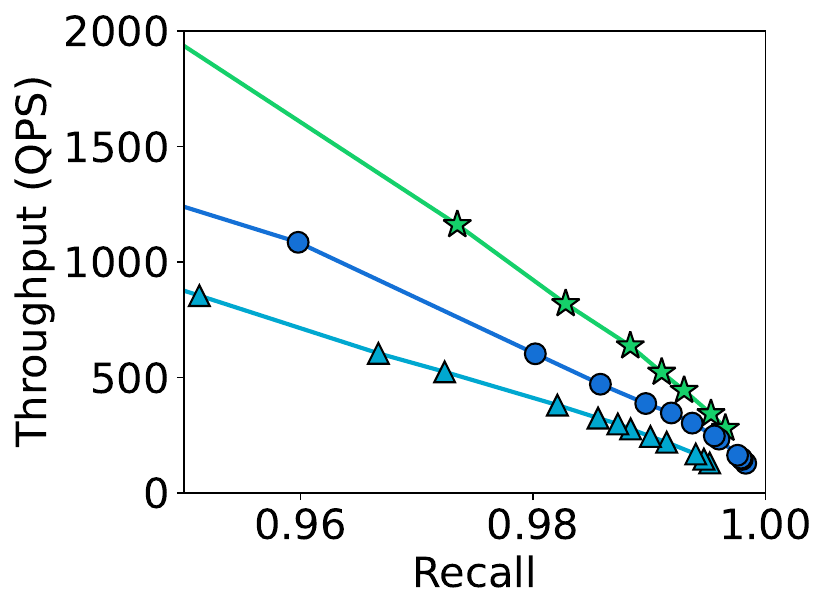}}
        \vspace{-0.3cm}
        \caption{OOD dataset.}
    \label{fig:ood}
\end{minipage}
\end{figure*}

\indexname{} supports other similarity metrics, for example, Euclidean distance. Figure~\ref{fig:l2} evaluates \nameindex{} against top baselines on two datasets, DPR-768 and GIST-960, without normalization and using Euclidean distance. \nameindex{} still outperforms the other baselines. Specifically, the difference between \nameindex{} and OPQIVFPQ\_RF demonstrates that the effectiveness of the learning techniques and search optimizations remains valid when using Euclidean distance.

\subsection{Further Discussion on Insertion/Deletion}

\begin{figure}[t]
\centering
\begin{minipage}[c]{0.49\columnwidth}
    \centering
    \includegraphics[width=\textwidth]{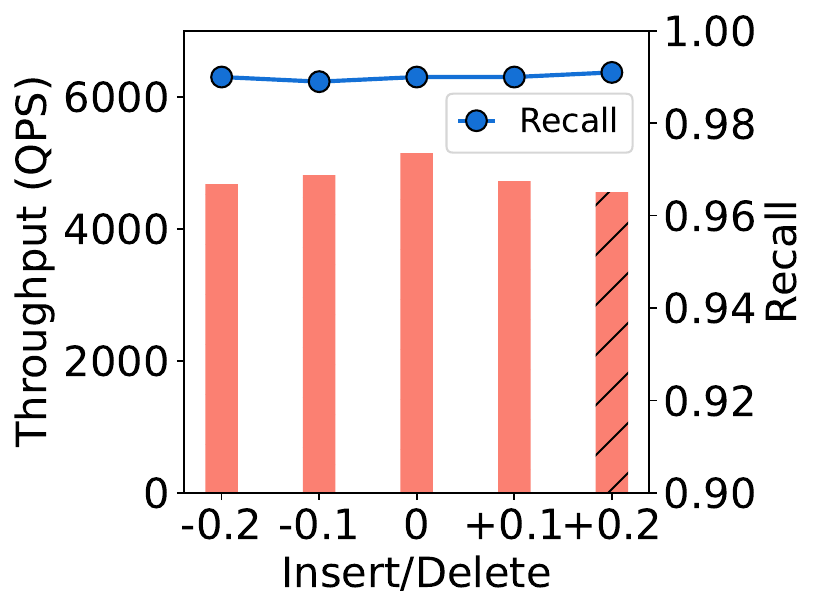}
    \caption{Search performance after insertion/deletion.}
    \label{fig:insdel}
\end{minipage}\hfill
\begin{minipage}[c]{0.480\columnwidth}
    \centering
    \includegraphics[width=\textwidth]{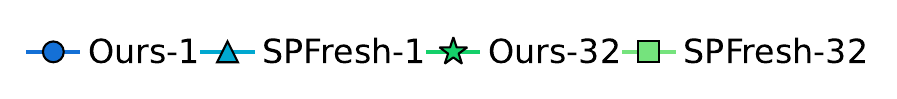}\\
    \includegraphics[width=0.90\textwidth]{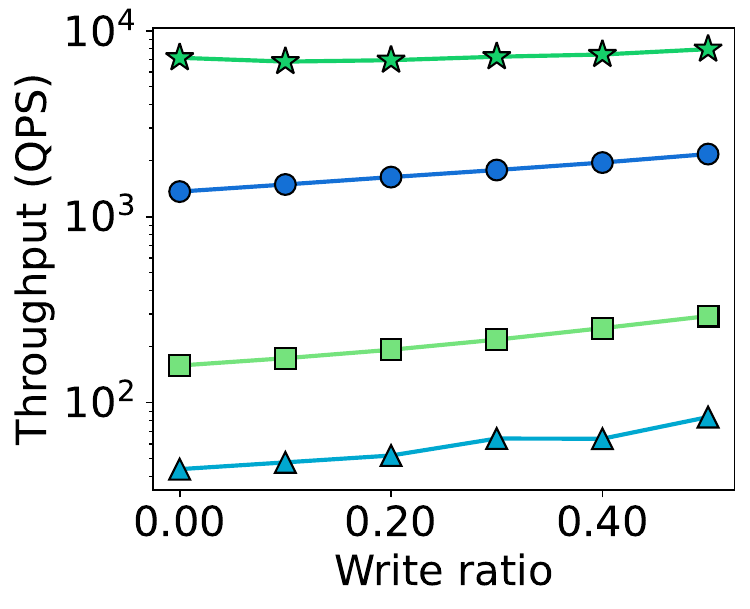}
    \vspace{-0.4cm}
    \caption{Comparison with SPFresh (with 1 and 32 threads).}
    \label{fig:spfresh}
\end{minipage}
\end{figure}

Both insertion and deletion are supported in \nameindex{} to minimize interference with search operations. For insertion, vectors are appended to the end of the memory buffer of the corresponding IVF partition. For deletion, tombstones for the deleted vectors are maintained. For a given search configuration $nprobe$ and $k'/k$, the search throughput is expected to decrease after a large number of insertions and deletions, because of the additional vectors to compare or tombstones to check during the filer stage. But, without changing the IVF partitioning, new nearest neighbors are likely to be found in the $nprobe$ partitions selected for a query, which maintains the recall. In Figure~\ref{fig:insdel}, we start with 1 million DPR-768 vectors in \nameindex{} and measure the search performance after 0.2 million deletions, 0.1 million deletions, 0.1 million insertions, and 0.2 million insertions. The same search configuration that achieved recall $\approx 0.99$ before any deletion or insertion is applied for all settings. We calculate the true nearest neighbors of query vectors after the insertions or deletions to report the recall, and we use 32 clients to measure the throughput. Compared to the original performance shown in the middle, the throughput of search decreases with more deletions or insertions, but the recall remains stable at around 0.99. As \nameindex{} is efficient to construct, it is recommended to rebuild the index when the size of dataset changed significantly by insertions and deletions.

SPFresh~\cite{spfresh} proposed a protocol to dynamically update IVF partitions as vectors are being inserted and deleted. By adjusting the partitions to limit the size of each partition, it aims to maintain predictable latency for each query. Under concurrent workload, it performs the update of IVF partitions via copy-on-write and manages a thread pool in the background to asynchronously handle splitting and merging of IVF partitions. We compare \nameindex{} against the SPFresh index under different concurrent workloads. SPFresh is designed for disk-based vector search, and thus we extend its IVF partition management to be in memory. Figure~\ref{fig:spfresh} shows that on the 1 million scale DPR-768 dataset at recall $\approx 0.97$, \nameindex achieves significantly higher throughput. Apart from the IVF update protocol, which introduces contention to the search query, SPFresh generates a large number of small partitions and indexes them using a graph index. For in-memory ANN search, \nameindex{} with its aggressive compression significantly reduces computation to perform high-recall search.

\subsection{Discussion on Out-of-Distribution Multi-Modal Dataset}

An emerging research problem in vector search is handling out-of-distribution (ODD) queries. In this setting, the distribution of query vectors differs from the distribution of data vectors in the database. This phenomenon is observed in multi-modal datasets where query and indexed vectors are from two respective modalities and embedded in different ways.
For example, queries are text embedding vectors, while the indexed vectors are image embedding vectors. The learning techniques in \nameindex{} can be adapted for the OOD setting. When preparing the training dataset, a sample query set is used and the approximate nearest neighbors of the query vectors are retrieved from the base index at step 3 in Figure~\ref{fig:build-b}. 

Figure~\ref{fig:ood} shows the throughput-recall trade-off on the Text-to-Image 10M dataset for \nameindex{}, the base index OPQIVFPQFS\_RF, and RoarGraph, a recent graph index built for OOD setting. The gap between \nameindex{} and OPQIVFPQ\_RF highlights the effect of our learning technique and search optimization. 
In particular, we use 100k out of 10 million sample queries available in that dataset for the training of search index parameters, and that improves the approximation of similarity score distribution by the compressed vectors. The search process then leverages the learned parameters to select more true nearest neighbors for the subsequent refine stage. 

The RoarGraph performance is evaluated locally based on the prebuilt index made available by the authors. It can be seen that RoarGraph achieves the highest performance. \nameindex{} can achieve competitive performance at very high recall.
Yet, the index construction of RoarGraph costs more compared to \nameindex{}, mainly due to the requirement for neighbor connections in the graph index, which leverages a large set of sample queries with their true nearest neighbors (10 million samples for their prebuilt index).
Based on experiments on our machine, it takes 3224s to build the index with the more efficient construction setting described in their paper~\cite{roargraph} with 1 million sample query vectors and their neighbors, excluding the cost to generate the true nearest neighbors.
In contrast, \nameindex{} consumes 716s for the index construction, including 336s for training set preparation and 82s for training the search index parameters. 
Since RoarGraph uses a graph index, it suffers from costly updates due to the reconstruction of connections in the neighborhood of inserted vectors, similar to HNSW and LVQ. 

We attribute the lower performance of \nameindex{} to the limitation of the space partitioning based on k-means clustering in the cross-modal datasets. As discussed in the RoarGraph paper, for the datasets involving queries of a different modality, the neighbors of a query are scattered over many partitions generated based on k-means, which increases the cost of scanning. Improving \nameindex{} for cross-modal settings requires a new space partitioning scheme, which presents a research challenge for future work. 

\subsection{Discussion on Low-Dimension Datasets}

\nameindex{} is designed for high-dimensional embedding vectors generated by deep learning models. By combining a filter stage using aggressively compressed vectors and a refine stage for reranking, it significantly reduces the computation complexity to achieve high-recall. On low-dimensional datasets, the search result is more sensitive to compression. We compare \nameindex{} with top baselines on two low-dimensional vector datasets used in the community.

Figure~\ref{fig:low-d} shows that on the Deep dataset with 256 dimensions, \nameindex{} is still the best performing index, whereas it falls behind some baselines on the SIFT dataset with 128 dimensions. 
Note that the vectors SIFT are not embedding vectors generated by a neural network. The value at each dimension is an 8-bit integer, and we cast it to float32 for the experiments.
Graph-based indexes (i.e., HNSW and LVQ) perform well on low-dimensional datasets, as we observe that they converge quickly to the neighborhood of a query during search. We also observed that ScaNN implementation in recent releases is more efficient than FAISS, especially on low-dimensional datasets, even when both scan the same ratio of space partitions for vectors compressed by 4-bit PQ.
On the SIFT dataset, the throughput-recall trade-off curve of \nameindex{} and OPQIVFPQ\_RF overlap with each other. It indicates that the learning technique provides little improvement for the low-dimensional vectors generated by the SIFT method.

\end{document}